\documentclass[amsmath,amssymb,showpacs,showkeywords,twocolumn]{revtex4}
\usepackage[dvips]{graphicx,color}
\usepackage{times}
\usepackage{mathrsfs}
\usepackage{amsmath}
\usepackage{graphicx}
\usepackage{epsfig}
\usepackage{dcolumn}
\usepackage{bm}
\begin{document}
\title{Current induced magnetization dynamics and magnetization switching in  
superconducting ferromagnetic hybrid (F$|$S$|$F) structures}
\author{Saumen Acharjee\footnote{saumenacharjee@gmail.com} and Umananda Dev 
Goswami\footnote{umananda2@gmail.com}}
\affiliation{Department of Physics, Dibrugarh University, Dibrugarh 786 004, 
Assam, India}

\begin{abstract}
We investigate the current induced magnetization dynamics and magnetization 
switching in an unconventional p-wave superconductor sandwiched between two 
misaligned ferromagnetic layers by numerically solving Landau-Lifshitz-Gilbert 
equation modified with current induced Slonczewski's spin torque term. A 
modified form of Ginzburg-Landau free energy functional has been used for this 
purpose. We demonstrated the possibility of current induced magnetization 
switching in the spin-triplet ferromagnetic superconducting hybrid structures 
with strong easy axis anisotropy and the condition for magnetization reversal. 
The switching time for such arrangement is calculated and is found to be 
highly dependent on the magnetic configuration along with the biasing current.
This study would be useful in designing practical superconducting-spintronic
devices.
\end{abstract}

\pacs{67.30.hj, 85.75.-d, 74.90.+n}

\maketitle

\section{Introduction}
During over last 15 years, a number of very interesting compounds have been 
discovered which reveal the coexistence of ferromagnetism and 
superconductivity in the same domain in bulk \cite{saxena,aoki,pfleiderer,huy,
flouquet,nandi}. The interplay between ferromagnetic order and 
superconductivity thus gains lots of attention from variety of research 
communities \cite{buzdin}. Among those, some peoples were hunting for 
superconductivity in a ferromagnetic spin valve made up of two ferromagnetic 
substances separated by a superconducting element (F$|$S$|$F system). In this
context it is to be noted that,
the spin triplet superconductivity in superconductor$|$ferromagnet (F$|$S) 
hybrid structures including F$|$S$|$F spin valves is a topic of intense 
research \cite{sangjun,tagirov,gu1,gu2,bergeret,moraru,zhu,leksin,banerjee} in 
the theoretical as well as experimental points of view for almost last two 
decades. 
The major interest of the F$|$S hybrid structures is due to the dissipation 
less flow of charge carriers offered by the superconducting environment. To 
completely understand this hybrid structure it is important to study the spin 
polarized transport. 

Moreover, the transport of spin is closely related to the phenomenon of 
current induced magnetization dynamics \cite{zutic} and spin transfer 
torque \cite{slonczewski,berger}. Spin transfer torque (STT), which is the 
building block of spintronics is based on the principle that, when a spin 
polarized current is applied into the ferromagnetic layers, spin angular 
momentum is transferred into the magnetic order. It is observed that for a 
sufficiently large current, magnetization switching can occur \cite{linder, 
linder1} in a magnetic layer. Thus, the flow of electrons can be served to 
manipulate the configuration of the spin valves. Traditionally, a lot of works 
had been done earlier on current induced magnetization dynamics and STT on 
ferromagnetic layers. Soon after, a lot attention have been made on 
anti-ferromagnetic layers \cite{nunez,tang,urazhdin,linder1} also. Making a 
hybrid structure of a superconductor with a ferromagnet and the concept of 
current induced magnetization dynamics suggest a very interesting venue for 
combining two different fields, namely superconductivity and spintronics 
\cite{linder2}. A few works had been done earlier on F$|$S hybrid 
structures \cite{waintal,takahashi,zhao,braude, linder3}. In 
Ref. \cite{linder}, supercurrent-induced magnetization 
dynamics in Josephson junction with two misaligned ferromagnetic layers had 
been studied and demonstrated the favourable condition for magnetization 
switching and reversal.

\begin{figure}[hbt]
\centerline
\centerline{ 
\includegraphics[width = 7.0cm, height = 5.0cm]{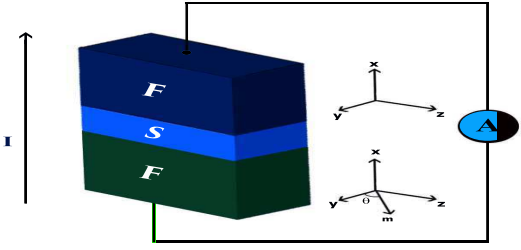}
}
\caption{The proposed experimental setup. An unconventional p-wave type 
superconductor is sandwiched in between two ferromagnetic layers. The 
magnetization orientation of the ferromagnetic layers are supposed to be 
misaligned by an angle $\theta$. When a current I is injected it gets 
polarized and transfer of spin torque to the magnetic order causing 
magnetization dynamics. Different colours of the ferromagnetic layers 
indicates the level of magnetization. Here, the bottom layer is hard in 
magnetization.}
\label{fig1}
\end{figure}

Motivated by the earlier works, in this paper we studied the current induced 
magnetization dynamics of a superconducting ferromagnet in a hybrid structure 
of F$|$S based on Landau-Lifshitz-Gilbert (LLG) equation with Slonczewski's 
torque (LLGS) using the Ginzburg-Landau-Gibb's free energy functional. The 
proposed experimental setup is shown in the Fig.\ref{fig1}, in which two 
ferromagnets are separated by a thin superconducting ferromagnet. The coercive 
fields of the ferromagnets are such that, the magnetization is hard in one 
layer while soft in the other and the  orientation of magnetization of the 
soft ferromagnetic layer is supposed to be misaligned with the hard 
ferromagnetic layers by an angle  $\theta$. When the junction is 
current-biased, it gets spin polarized in the hard layer and thus transfer 
angular momentum to the magnetic order. This generates an induced 
magnetization contributing to the magnetic order. The dynamics of this induced 
magnetization has been studied by numerically solving the LLGS equation. 

The paper is organized as follows. In the Section II, a theoretical framework 
of the proposed setup is developed. The results of our work is discussed in 
the Section III  by solving LLGS equation numerically. Finally we conclude 
our work in the Section IV.

\section{Theory}
To study the current induced magnetization dynamics of a ferromagnetic 
superconductor with easy axis anisotropy in F$|$S$|$F spin valve, we utilized 
Landau-Lifshitz-Gilbert (LLG) equation with the Slonczewski's spin transfer 
torque (LLGS). The resulting LLGS equation takes the form
\begin{equation}
\label{eq1}
\frac{\partial  \textbf{M}}{\partial t}  = -\gamma (\textbf{M} \times \textbf{H}_{eff}) + {\alpha} (\textbf{M} \times \frac{\partial\textbf {M}}{\partial t})+ \textbf{T},
\end{equation}
where $\gamma$ is the gyromagnetic ratio,  $\alpha$ is the Gilbert's damping 
constant and $H_{eff}$ is the effective magnetic field of superconducting 
ferromagnet. $\textbf{T}$ is the current induced spin transfer torque and can 
be read as \cite{linder}

\begin{equation}
\label{eq2}
\textbf{T} = I\zeta(\textbf{M}\times[\textbf{M}\times(\textbf{M}_{T}-{\epsilon}\textbf{M}_{B}]), 
\end{equation}
where $\textbf{M}_T$ and $\textbf{M}_B$ respectively represents the normalized 
magnetization vector in the top and bottom magnetic layers of the spin valve and
is taken as $\textbf{M}_{T} = (0,1,0)$ and $\textbf{M}_{B} = (0,\cos\theta,\sin\theta)$ such that for $\theta = 0$, the configuration is parallel and is 
anti-parallel for $\theta = \pi$. $\epsilon$ provides the factor of asymmetry 
in polarization in top and bottom ferromagnetic layers. The term $\zeta$ is
given by

\begin{equation}
\label{eq2a}
\zeta = \frac{\nu\hbar\mu_0}{2e m_0V}.
\end{equation}
Here, $e$ is the electronic charge, $\nu$ is the polarization efficiency, 
$\hbar$ is the Planck's constant, $\mu_0$ is the magnetic permeability, $m_0$ is
the amplitude of magnetization and $V$ is the volume of the system. $I$ is the
applied current bias. The effective magnetic field of the system can be 
obtained from the functional derivative of the free energy with respect to the 
magnetization: 
\begin{equation}
\label{eq3}
\textbf{H}_{eff} = -\frac{d F}{d \textbf{M}}. 
\end{equation}
The free energy functional $F(\psi,\textbf{M})$ can be given by \cite{shopova}
\begin{equation}
\label{3a}
F(\psi,\textbf{M}) = \int d^3\textbf{r} f(\psi,\textbf{M}),
\end{equation}
where $f(\psi,\textbf{M})$ gives the free energy density of a spin-triplet 
superconductor and can be read as \cite{dahl,shopova}
\begin{equation}
\label{eq4}
f ({\psi,\textbf{M}}) = f_S({\psi}) + f_F({\textbf{M}}) + f_{int} (\psi,\textbf{M}) + \frac{\textbf{B}^2}{8\pi} - \textbf{B}.\textbf{M},
\end{equation}
where $\psi\; (\equiv \psi_j; j = 1,2,3)$ is the superconducting order 
parameter and is a three dimensional complex vector, $\textbf{M}$ is the 
magnetization vector, which characterizes the ferromagnetism, $f_S({\psi})$ 
gives the 
superconductivity, while the ferromagnetic order is described by 
$f_F(\textbf{M})$. The interaction of the two orders, $\textbf{M}$ and 
$\psi$ is described by the term $f_{int}(\psi,\textbf{M})$. The last two terms 
in equation (\ref{eq4}) account the contribution of magnetic energy on free 
energy with magnetic induction  $\textbf{B} = \textbf{H}+4\pi\textbf{M} 
= \nabla \times \textbf{A}$.

The superconductivity of the system is described by the term $f_S(\psi)$ under 
the condition $\textbf{H} = 0$ and $\textbf{M} = 0$ and can be written as 
\cite{dahl,shopova, shopova-n}
\begin{multline}
\label{eq4a}
f_S(\psi) = f_{grad}(\psi)+a_s|\psi|^2+\frac{b_s}{2}|\psi|^4 \\+ \frac{u_s}{2}|\psi^2|^2+\frac{v_s}{2}\sum_{i=1}^{3}|\psi|^4,
\end{multline}
where $f_{grad}$ can be written as \cite{dahl}
\begin{multline}
f_{grad} = K_1(D_i\psi_j)^*(D_i\psi_j)+K_2[(D_i\psi_i)^*(D_j\psi_j)
\\+(D_i\psi_j)^*(D_j\psi_i)]+K_3(D_i\psi_i)^*(D_i\psi_i)
\end{multline}
with $D_i = -i\hbar(\frac{\partial}{\partial{x_i}})+ 2\frac{|e|}{c}A_i$ being 
the covariant derivative, $u_s$ describes the anisotropy of the spin triplet 
Cooper pair and the crystal anisotropy is described by $v_s$. $a_s$ and $b_s$ 
are positive material parameters. The term $f_{F}(\textbf{M})$ in (\ref{eq4}) 
describes the ferromagnetic ordering of the material and is given by \cite{dahl,shopova}

\begin{equation}
\label{eq4b}
f_F(\textbf{M}) = c_f\sum_{j=1}^3|\nabla_j\textbf{M}_j|^2 + a_f\textbf{M}^2 + \frac{b_f}{2}\textbf{M}^4.
\end{equation}
While the term $f_{int}(\psi,\textbf{M})$  in (\ref{eq4}) corresponds to the 
interaction of ferromagnetic order with the complex superconducting order and 
can be written as 

\begin{equation}
\label{eq4c}
f_{int}(\psi,\textbf{M}) = i\gamma_0\textbf{M}.(\psi \times \psi^*) +\delta\textbf{M}^2|\psi|^2, 
\end{equation}
where $\gamma_0$ term provides the superconductivity due to ferromagnetic order,
while $\delta$ term makes the model more realistic as it represent the strong 
coupling and can be both positive and negative values. Rewriting the free 
energy $f(\psi,\textbf{M})$ in a dimensionless form by redefining the order 
parameters $\psi_j = b_s^{-\frac{1}{4}}\phi_je^{i\theta_j}$ and  
$\textbf{m} = b_f^{-\frac{1}{4}}\textbf{M}$, the free energy (\ref{eq4}) takes 
the form
\begin{multline}
\label{eq4d}
f = f_{grad} + r\phi^2 + \frac{1}{2}\phi^4- 2t_1[\phi_1^2\phi_2^2\sin^2(\theta_2-\theta_1)
\\+ \phi_1^2\phi_3^2\sin^2(\theta_1-\theta_3) + \phi_2^2\phi_3^2\sin^2(\theta_2-\theta_3)]
\\- v[\phi_1^2\phi_2^2 
 + \phi_2^2\phi_3^2 + \phi_3^2\phi_1^2] + w\textbf{m}^2 + \frac{1}{2}\textbf{m}^4
\\+ 2\gamma_1\phi_1\phi_3\textbf{m}\sin(\theta_3-\theta_1) + \gamma_2\phi^2\textbf{m}^2 
-v_1\textbf{B}.\textbf{m},
\end{multline}
where the parameters, $r = \frac{a_s}{b^{\frac{1}{2}}}$ , 
$w = \frac{a_f}{b_f^{\frac{1}{2}}}$ , $t_1 = \frac{u_s}{b}$, 
$v = \frac{v_s}{b}$,
\\ $\gamma_1 = \frac{\gamma_0}{b^{\frac{1}{2}}b_f^{\frac{1}{4}}}$,
 $\gamma_2 = \frac{\delta}{(bb_f)^{\frac{1}{2}}}$ and $v_1 = b_f^{\frac{1}{4}}$ with $b = (b_s + u_s + v_s).$

The coexistence of superconductivity and ferromagnetism was first observed in 
UGe$_2$ \cite{saxena, bespalov}  within a limited pressure range 
(1.0 - 1.6 GPa). In following years, same coexistence was found in 
URhGe \cite{aoki, bespalov} and UCoGe \cite{bespalov, huy, bastien} at ambient 
pressure, and in UIr \cite{akazawa} similar to the case of UGe$_2$, i.e. 
within a 
limited pressure range (2.6 - 2.7 GPa). These Uranium-based (U-based) compounds,
with the coexistence of ferromagnetism and superconductivity, exhibit 
unconventional properties of ground state in a strongly correlated 
ferromagnetic system. One of the interesting features of these U-based 
ferromagnetic superconductors is that, this type of superconductivity was found 
to occur within the vicinity of a quantum critical point (QCP). The 
critical pressure, or critical chemical composition is referred to as the
QCP, where the ordering temperature is tuned to $T_C = 0$ K. It should be noted that in general, the U-based ferromagnetic superconductors 
have a very strong easy-axis magneto crystalline anisotropy 
\cite{shopova, bespalov}. 
However, the free energy in equation (\ref{eq4d}) is isotropic. To account the 
contribution of anisotropy in free energy, we introduce a term $K_{an}$ 
\cite{linder} 
resulting in an effective field of the form $H_{an} =  (K_{an}m_y/M_0)\hat{y}$.
Here we direct the anisotropy axis in parallel to the y-direction and 
contribution along the anisotropy axis is being considered. In view of this, 
the LLGS equation (\ref{eq1}) takes the form  
\begin{multline}
\label{eq5}
\frac{\partial  \textbf{m}}{\partial t}  = -\gamma [\textbf{m} \times (2w\textbf{m} + 2\textbf{m}^3 + 2\gamma_1\phi_1\phi_3\sin(\theta_3 - \theta_1)\hat{y} 
\\+ 2\gamma_2\phi^2\textbf{m} -v_1\textbf{B} + \frac{K_{an}m_y}{M_0}\hat{y})]+ {\alpha} (\textbf{m} \times 
\frac{\partial\textbf {m}}{\partial t}) + \textbf{T},
\end{multline}
where $\textbf{m}_y$ is the component of \textbf{m} along the anisotropy axis 
which we direct parallel to y-axis with $\textbf{B} = 
-B_{0}\hat{z}$. The equation (\ref{eq5}) is a non-linear coupled differential 
equation in $\textbf{m}$ and can be transformed into the following form
\begin{multline}
\label{eq6a}
\frac{dm_x}{d\tau} = \alpha\epsilon I \sin\theta\; m_y^3(\tau) + \alpha I (1-\epsilon \cos\theta)m_y^2(\tau)m_z(\tau) 
\\+ \alpha I (1-\epsilon \cos\theta)m_z^3 + \alpha I m_x^2(\tau)[\epsilon \sin\theta\; m_y(\tau)
\\ + (1-\epsilon \cos\theta)\;m_z(\tau)] + m_y(\tau)[-B_0 v_1 + m_z(\tau)(K_{an} 
\\+ 2w + \alpha \epsilon I \sin\theta\; m_z(\tau) + 2\phi^2\gamma_2)] + 2\sin\beta\; m_z(\tau)\gamma_1 \phi_1 \phi_3
\\- m_x(\tau)[m_z(\tau)(\epsilon I \sin\theta + \alpha B_0 v_1) + \alpha m_y^2(\tau)(K_{an} + 2w 
\\+ 2\phi^2\gamma_2) + m_y(\tau)(-I + \epsilon I \cos\theta + 2\alpha \gamma_1 \phi_1\phi_3 \sin\beta )]
\\/ [1 + \alpha^2(m_x^2(\tau) + m_y^2(\tau)+ m_z^2(\tau))],
\end{multline}
\begin{multline}
\label{eq6b}
\frac{dm_y}{d\tau} =  -\alpha\epsilon I \sin\theta\; m_x^3(\tau) + m_x(\tau)[-\alpha\epsilon I \sin\theta(m_x^2(\tau) 
\\+ m_z^2(\tau))+ B_0v_1]+m_x^2(\tau)[-I+\epsilon I \cos\theta +\alpha m_y(\tau)(K_{an} 
\\+ 2w  + 2\phi^2\gamma_2) + 2\alpha \sin\beta\;\gamma_1\phi_1\phi_3]+ m_z(\tau)[m_y(\tau)(-\epsilon I \sin\theta  
\\- \alpha B_0 v_1 + \alpha m_z(\tau)(K_{an} + 2w + 2\phi^2\gamma_2)) + m_z(\tau)(I(-1 
\\+ \epsilon \cos\theta) + 2\alpha \sin\beta\; \gamma_1 \phi_1\phi_3)]/[ 1+ \alpha^2(m_x^2(\tau) 
\\+ m_y^2(\tau)+ m_z^2(\tau))],
\end{multline}
\begin{multline}
\label{eq6c}
\frac{dm_z}{d\tau}= \alpha I (1+\epsilon \cos\theta)m_x^3(\tau) + m_x^2(\tau)(I\epsilon \sin\theta + \alpha B_0 v_1)
\\ - m_x(\tau)[\alpha I (1-\epsilon \cos\theta)m_y^2(\tau) + \alpha I (1-\epsilon \cos\theta)m_z^2(\tau) 
\\ + m_y(\tau)(K_{an} + 2w + 2\phi^2\gamma_2) + 2\sin\beta\; \gamma_1 \phi_1 \phi_3] 
\\ + m_y(\tau)[m_y(\tau)(\epsilon I \sin\theta + \alpha B_0 v_1 - \alpha m_z(\tau)(K_{an} + 2w 
\\+ 2\phi^2\gamma_2)) + m_z(\tau)(I - \epsilon I \cos\theta - 2\alpha \sin\beta \gamma_1 \phi_1\phi_3)]
\\/ [1 + \alpha^2(m_x^2(\tau) + m_y^2(\tau) + m_z^2(\tau))],
\end{multline}
where $\beta = (\theta_3 - \theta_1)$ represents the phase mismatch of 
surviving components of the superconducting order parameter. For a
realistic situation this phase mismatch should not be very large and hence we
have taken the $\beta$ to be equal to 0.1$\pi$ arbitrarily to have a similarity
with the practical situation. As U-based 
ferromagnetic superconductors have a very strong magneto crystalline anisotropy
\cite{shopova,bespalov}, to model a realistic superconducting ferromagnet,  
the anisotropy field can be taken as \cite{bespalov} $K_{an} \sim 10^3 $, the 
asymmetry factor is taken as $\epsilon = 0.1$ with magnetic induction 
$B_0 = 0.1$ and $\zeta = 1$. Furthermore, we have set \cite{shopova} 
$v_1 = w = 0.1$, $\phi_1 = \phi_3 = \frac{\phi}{\sqrt{2}}$  and initially 
$\gamma_1 = 2\gamma_2 = 0.2, $ which make F$|$S$|$F spin valve system more 
realizable. 

\section{Results and Discussions}
\begin{figure*}[hbt]
\centerline
\centerline{
\includegraphics[scale=0.5]{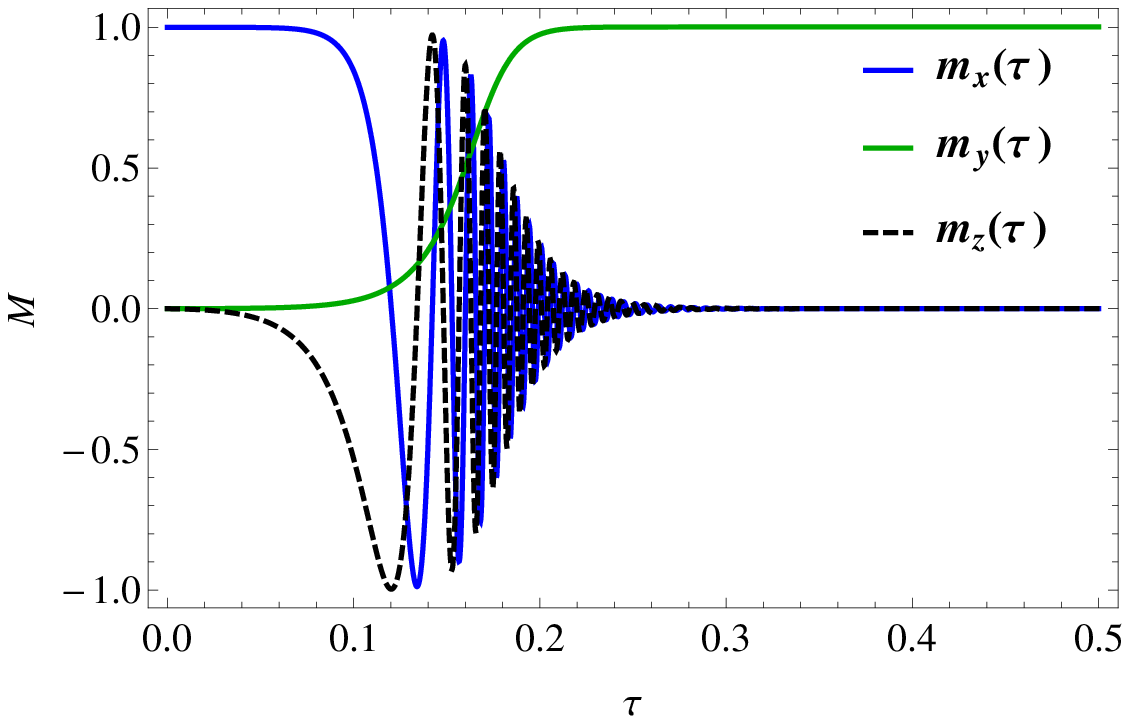}
\hspace{0.5cm}
\includegraphics[scale=0.5]{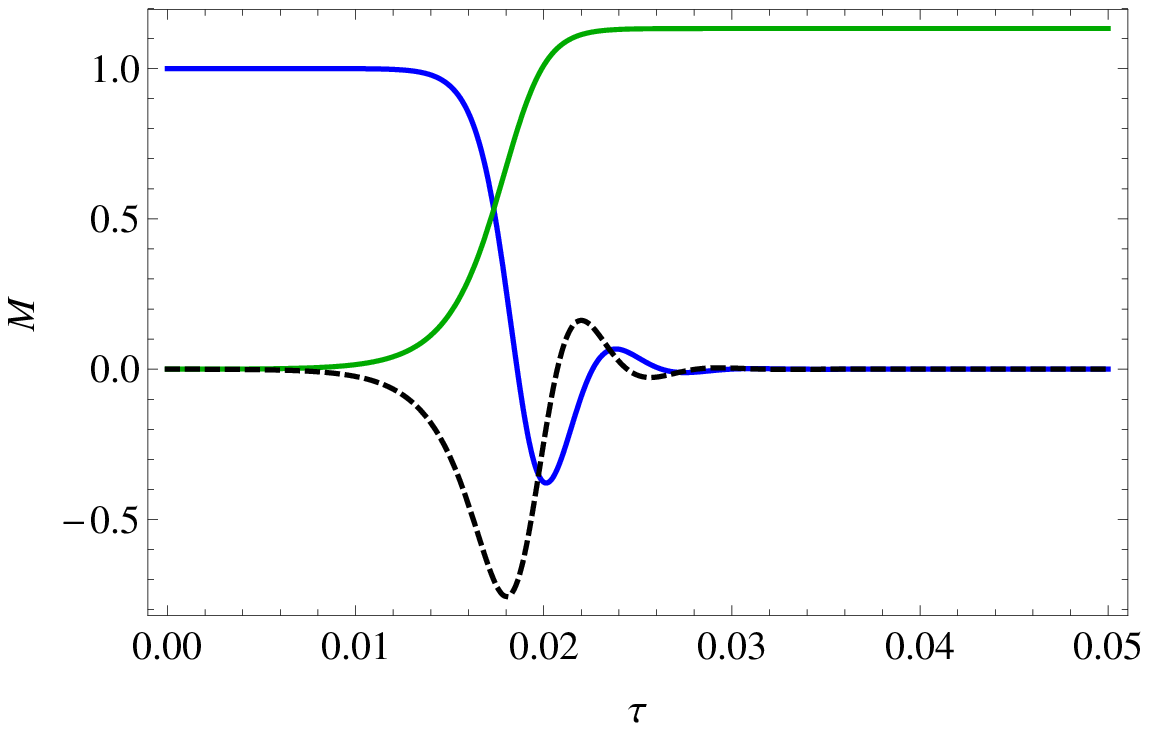}
\vspace{0.3cm}}
\centerline{\hspace{-0.0cm}
\includegraphics[scale=0.475]{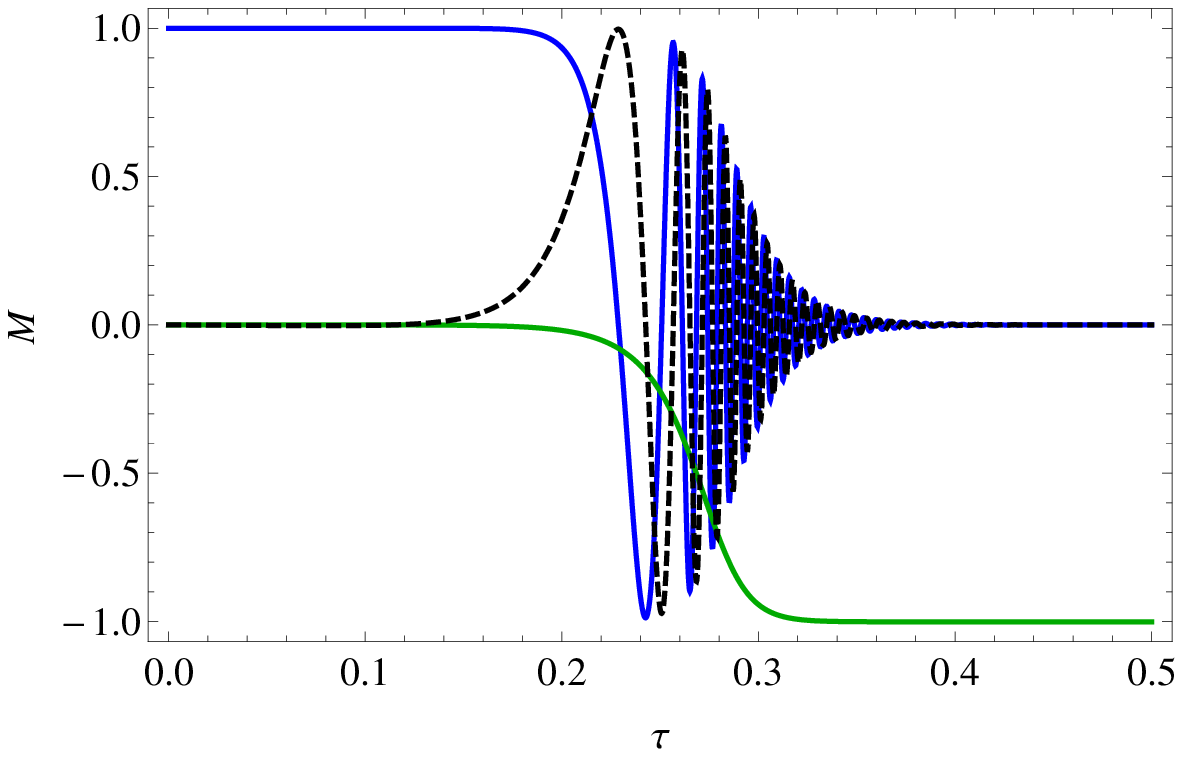}
\hspace{0.5cm}
\includegraphics[scale=0.475]{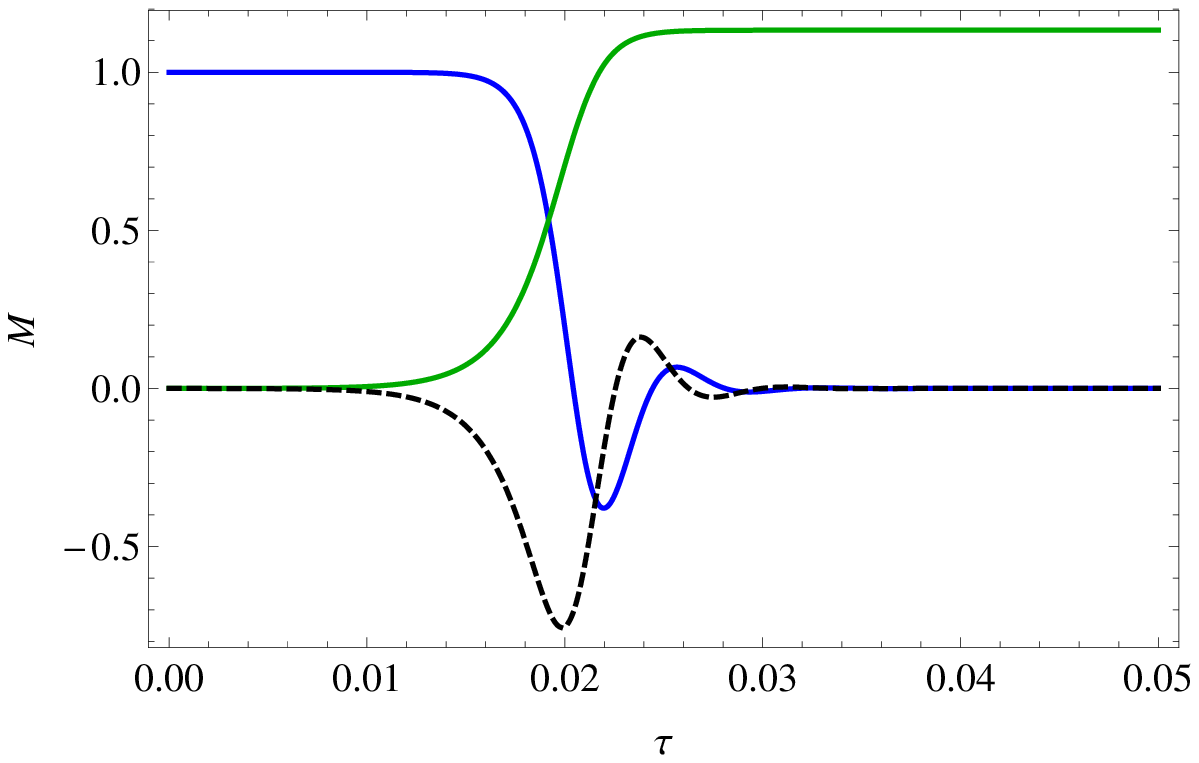}
}
\centerline{
\includegraphics[scale=0.45]{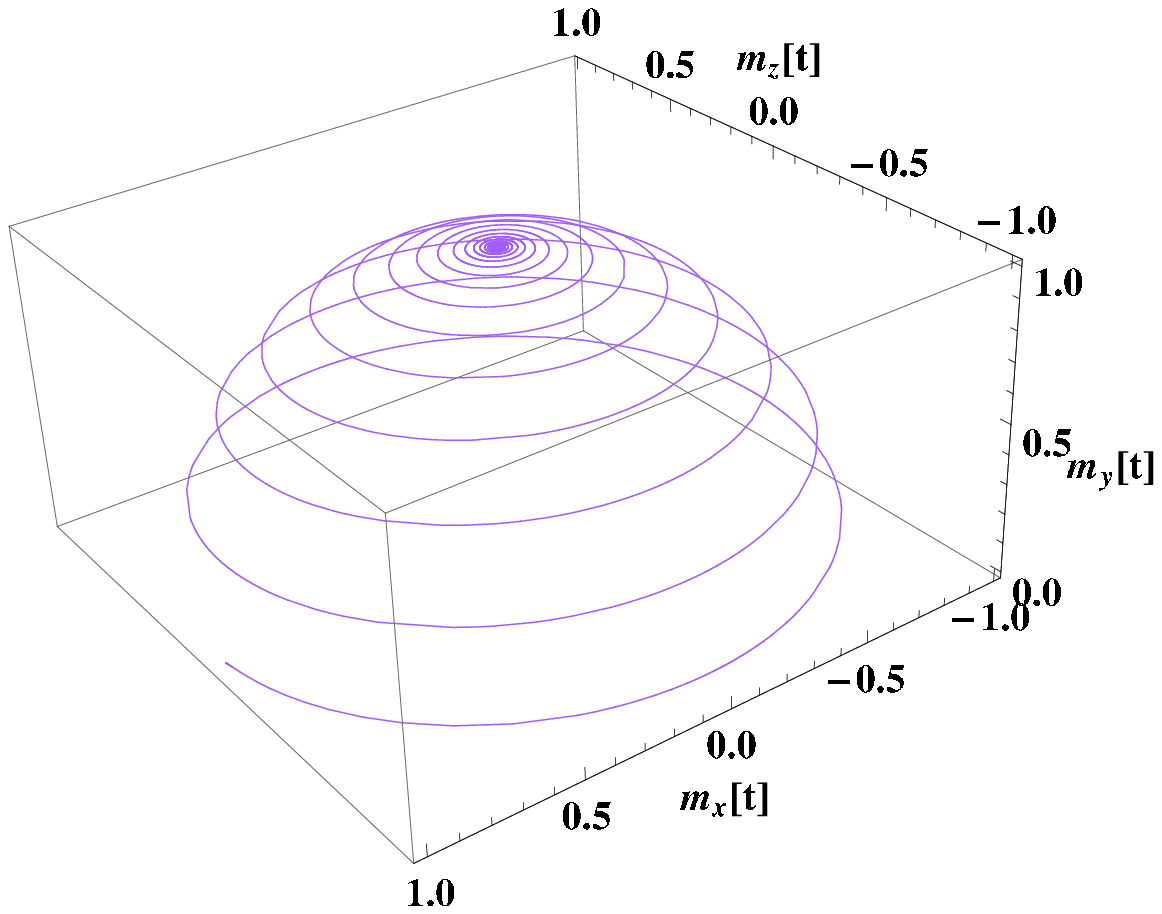}
\hspace{0.5cm}
\includegraphics[scale=0.47]{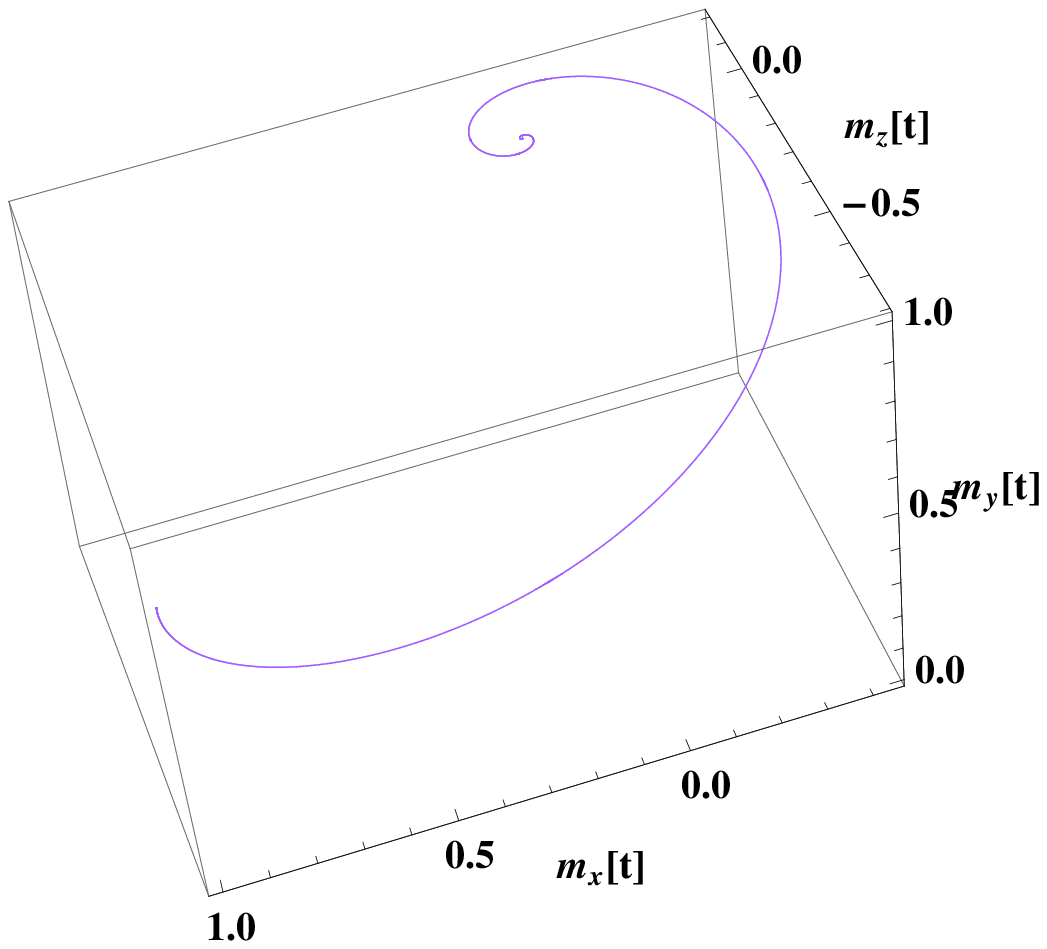}
\vspace{0.3cm}}
\centerline{\hspace{-0.0cm}
\includegraphics[scale=0.45]{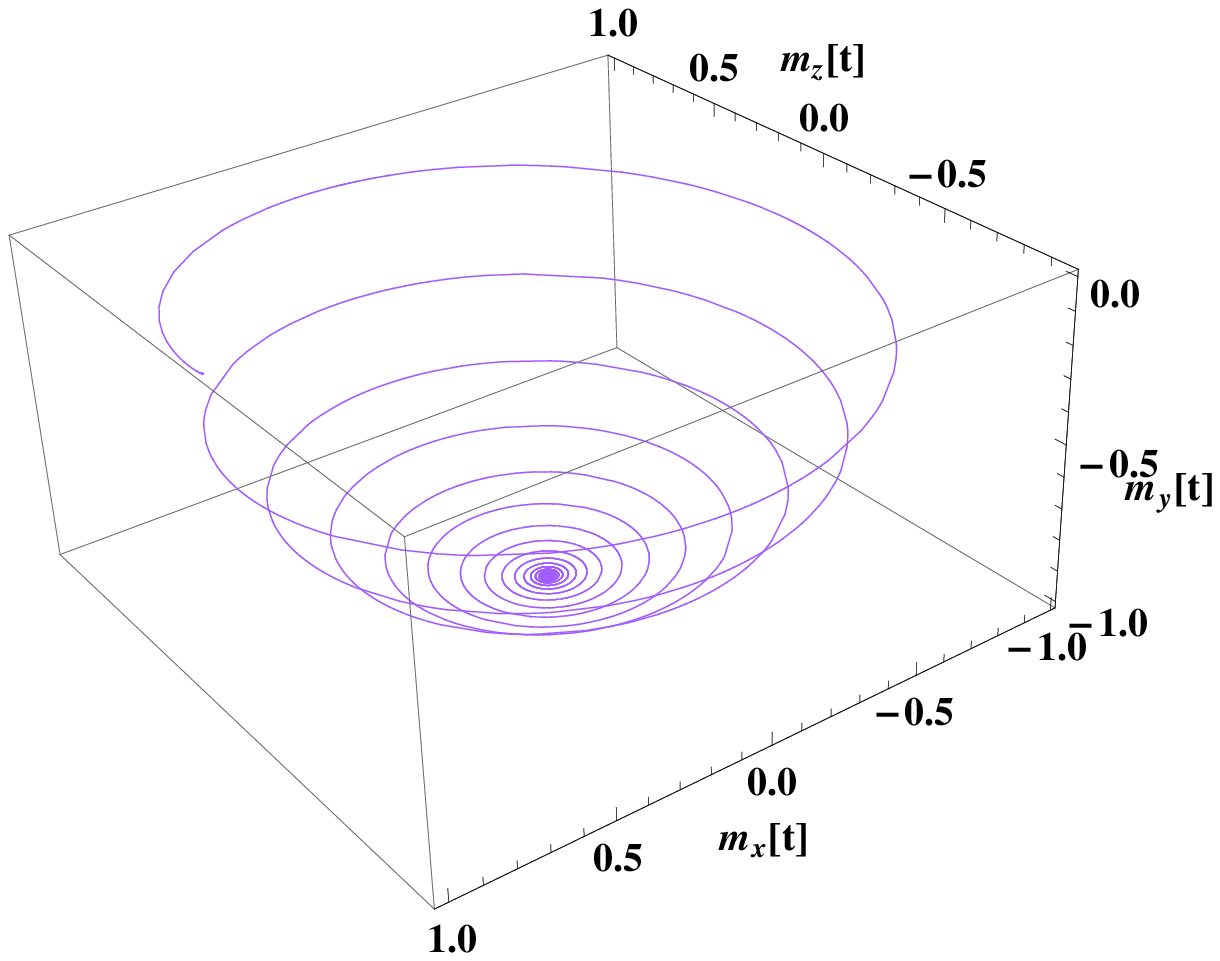}
\hspace{0.5cm}
\includegraphics[scale=0.45]{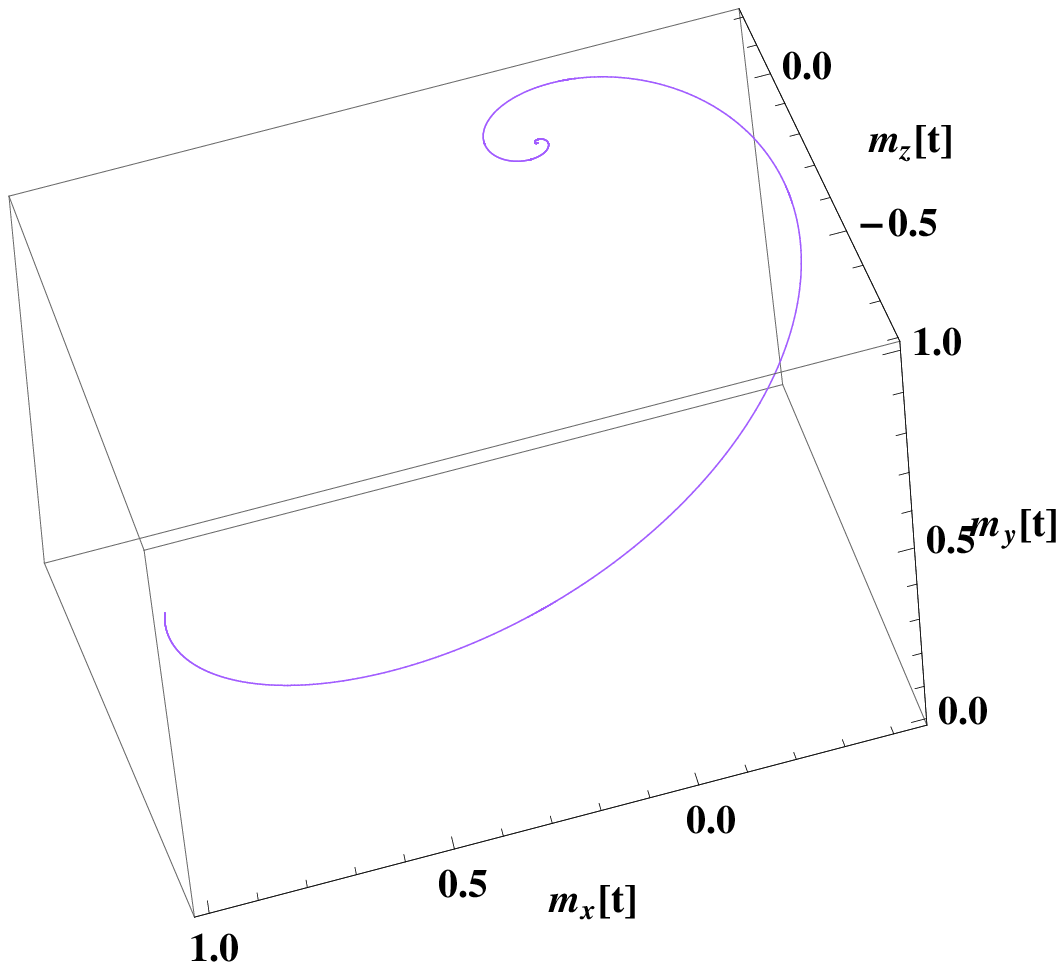}
}
\caption{(First four plots) The time evolution of the normalized components of magnetization with 
initial angle of misalignment $\theta = 0.1 \pi$ and with $\gamma_1 = 2\gamma_2
= 0.2.$ The plots in the left panels depicted for the weak damping 
$\alpha = 0.05$, while the plots in the right panels are for strong damping 
$\alpha = 0.5$ with a current biasing of $0.1$ mA and $0.252$ mA respectively 
from top to bottom panels. The corresponding parametric 
graphs representing the behaviour of the magnetization are displayed in the
last four plots.}
\label{fig2}
\end{figure*}

\begin{figure*}[hbt]
\centerline
\centerline{
\includegraphics[scale=0.46]{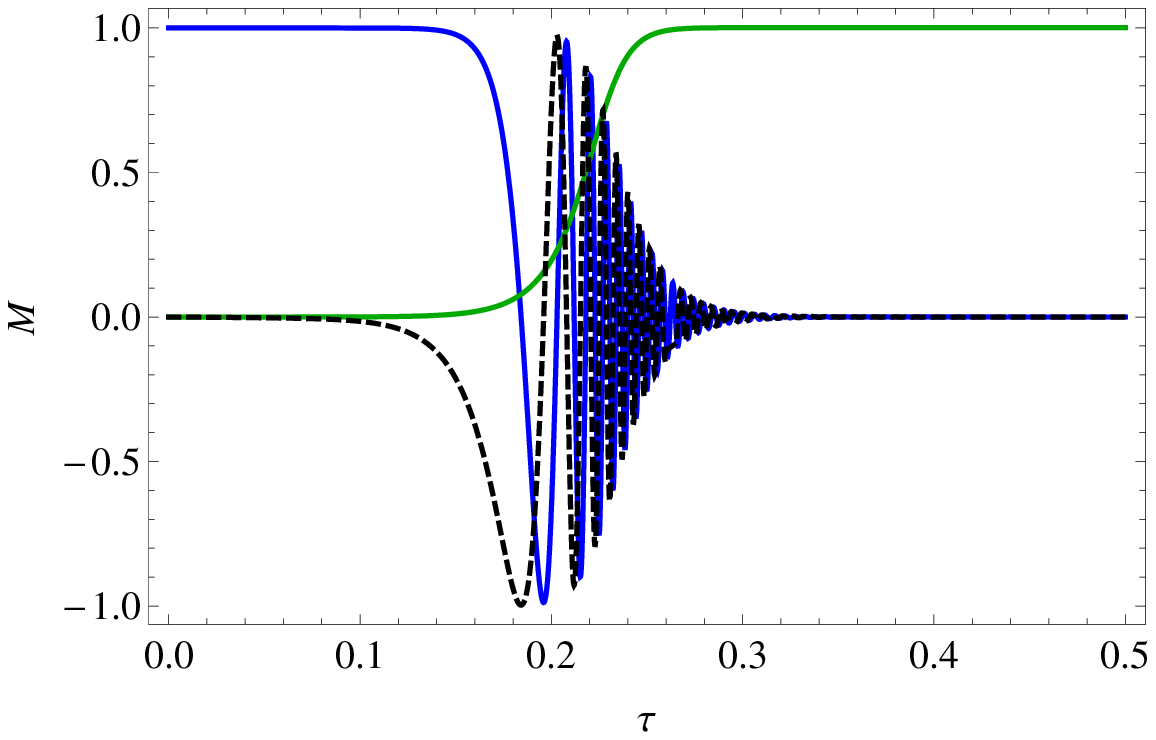}
\hspace{0.5cm}
\includegraphics[scale=0.46]{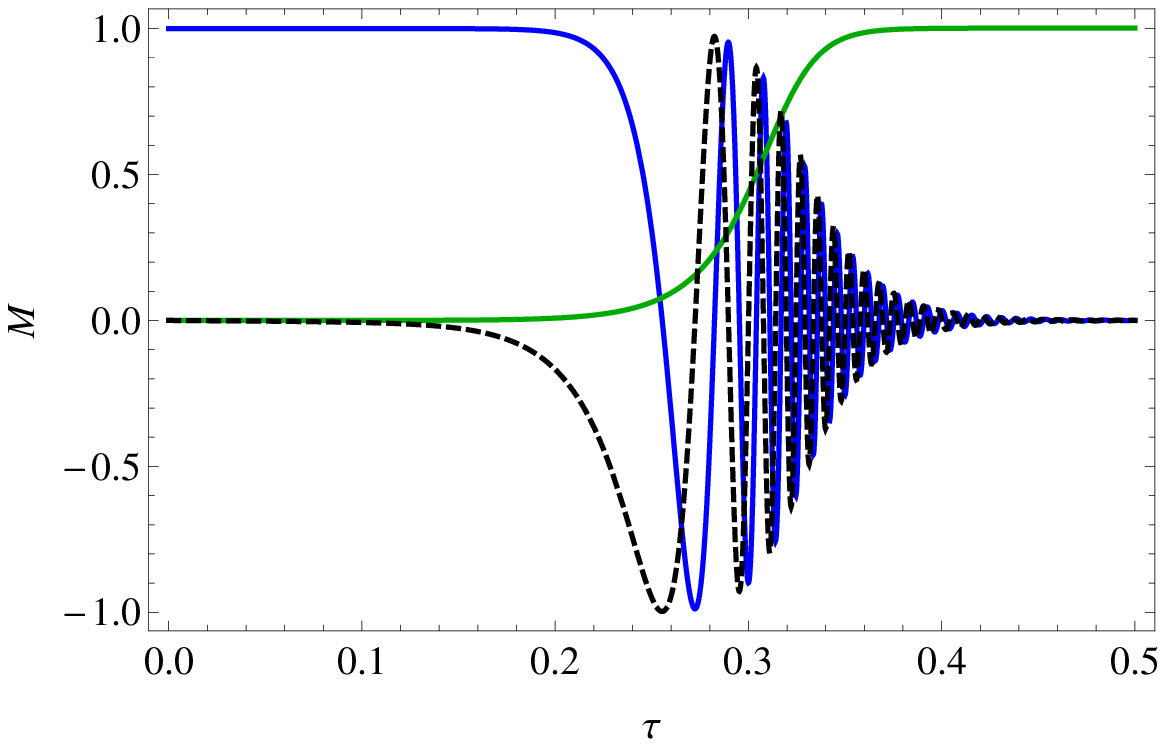}
\vspace{0.3cm}}

\centerline{\hspace{-0.0cm}
\includegraphics[scale=0.45]{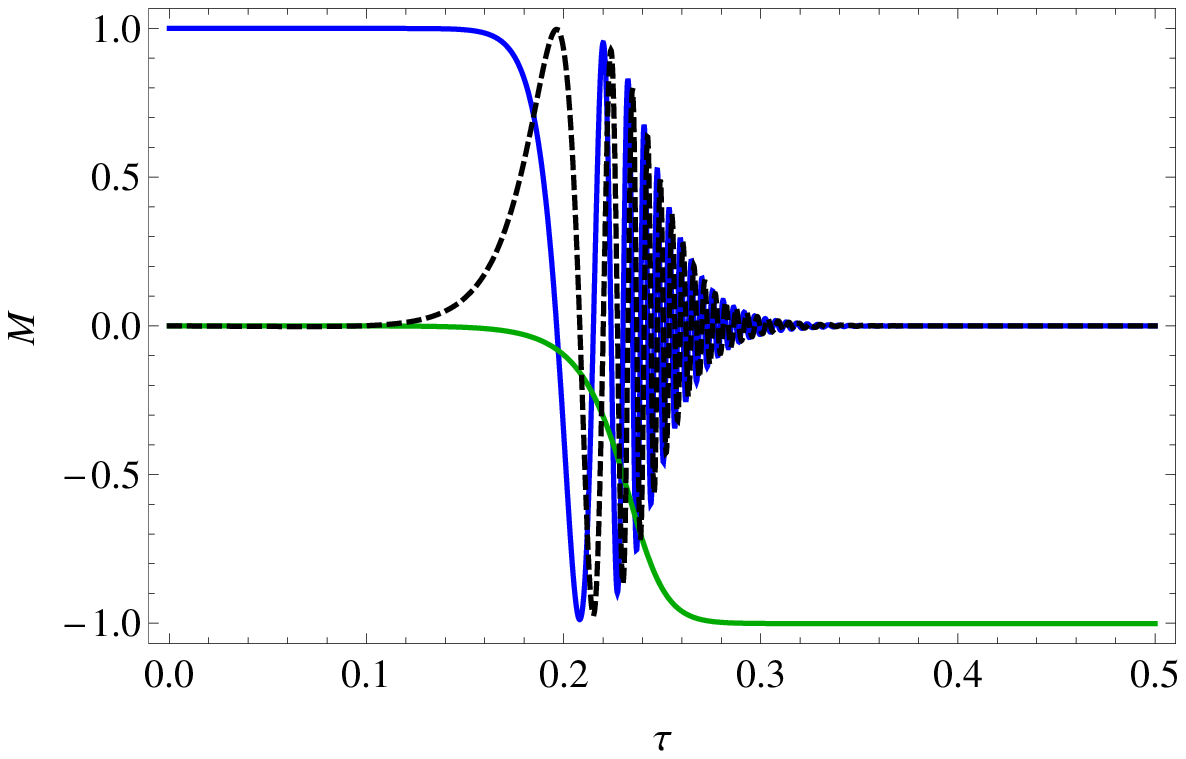}
\hspace{0.5cm}
\includegraphics[scale=0.45]{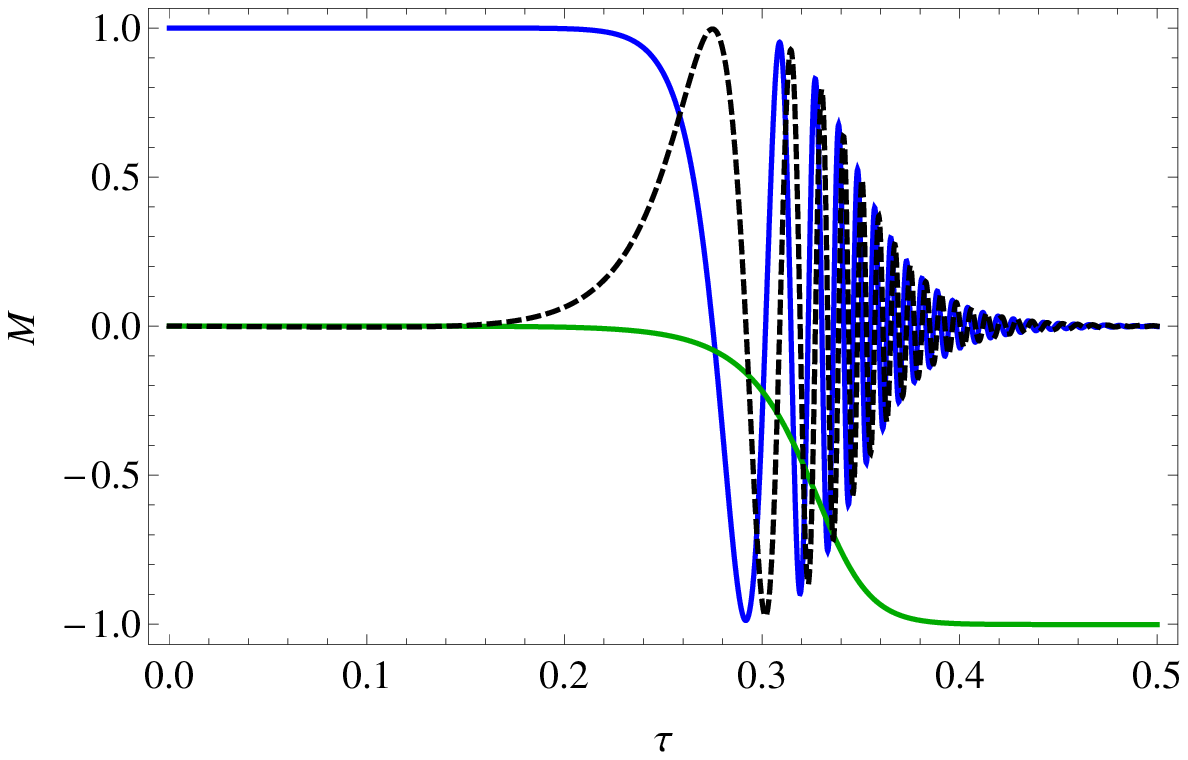}
}
\centerline{
\includegraphics[scale=0.45]{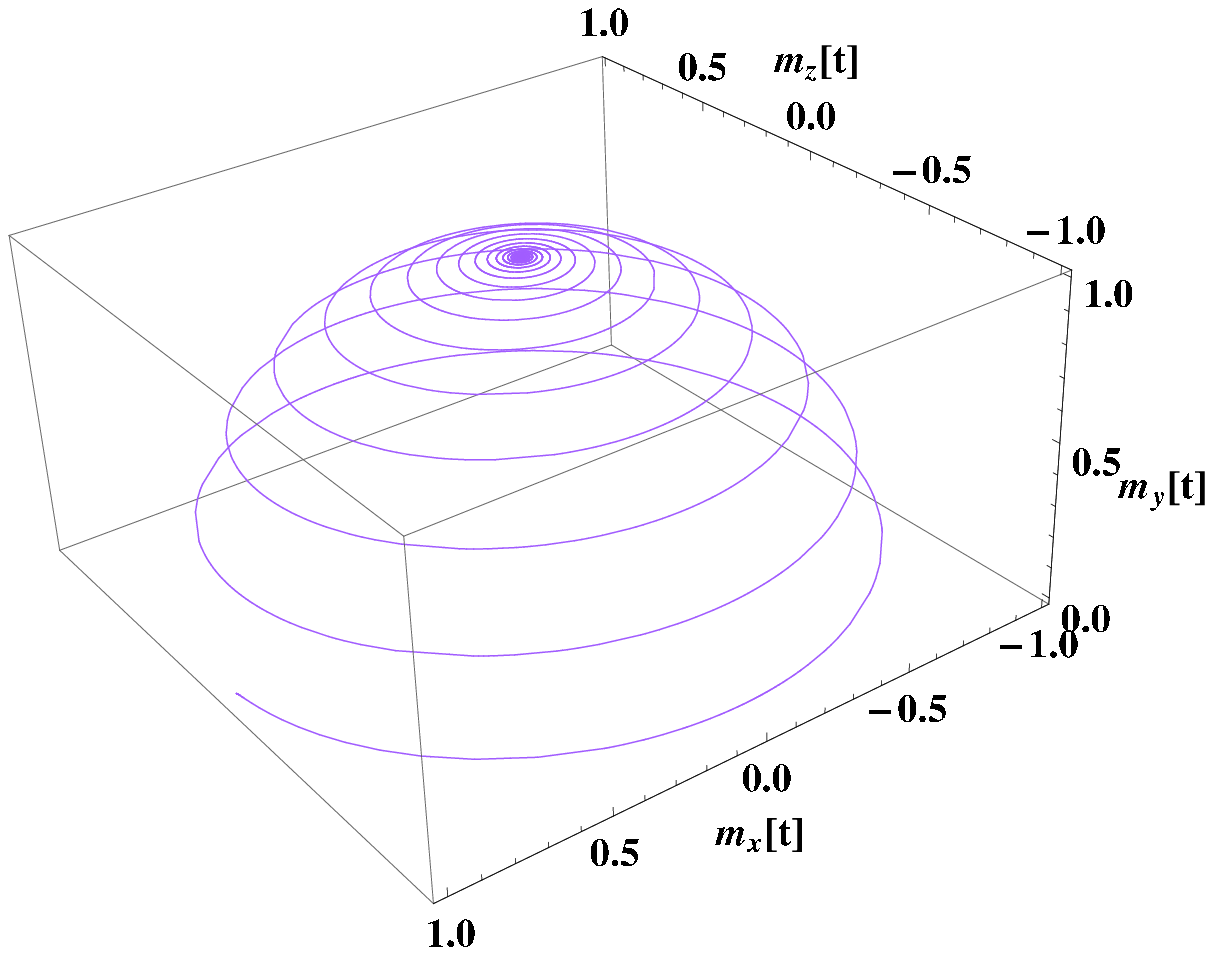}
\hspace{0.5cm}
\includegraphics[scale=0.45]{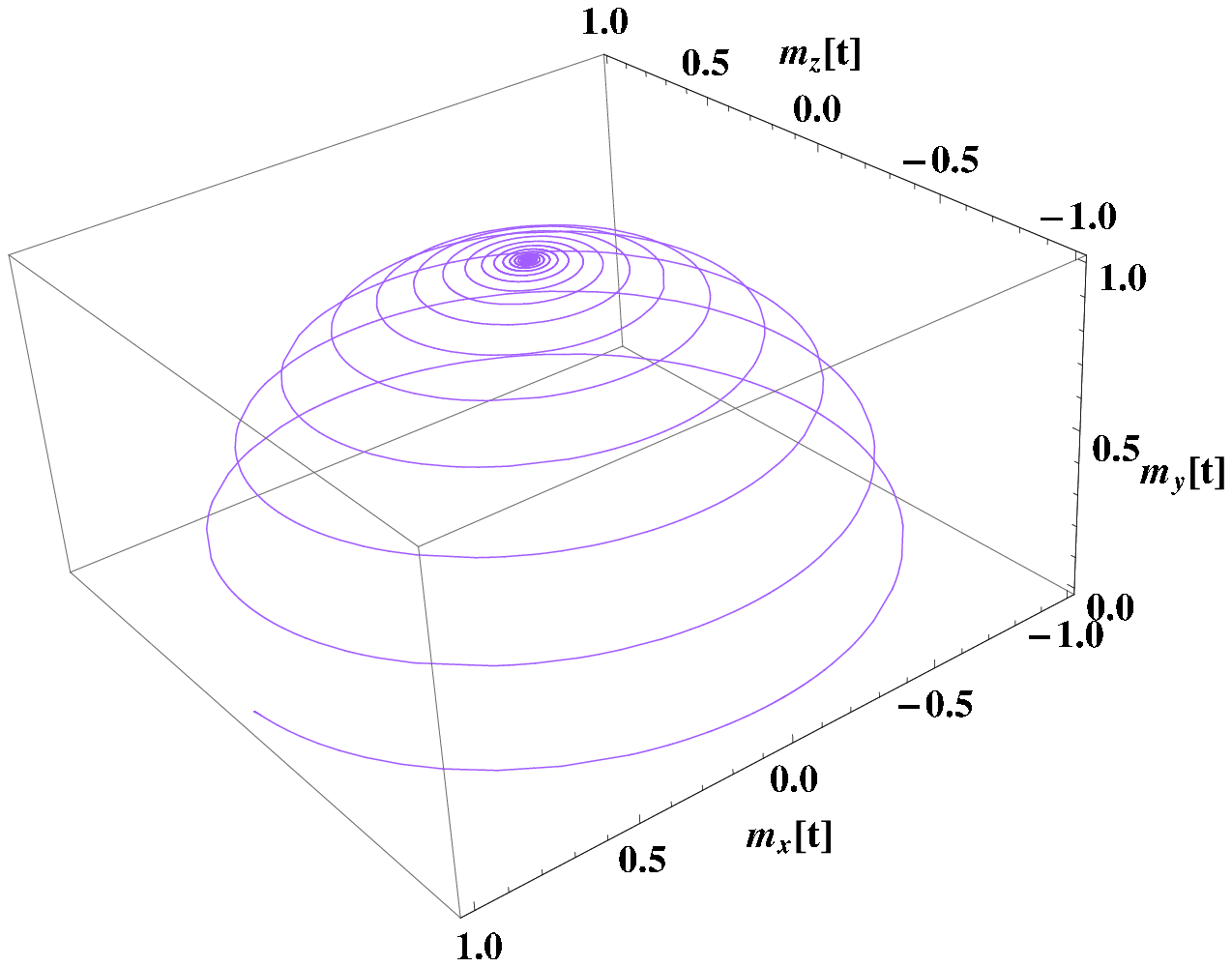}
\vspace{0.3cm}}

\centerline{\hspace{-0.0cm}
\includegraphics[scale=0.45]{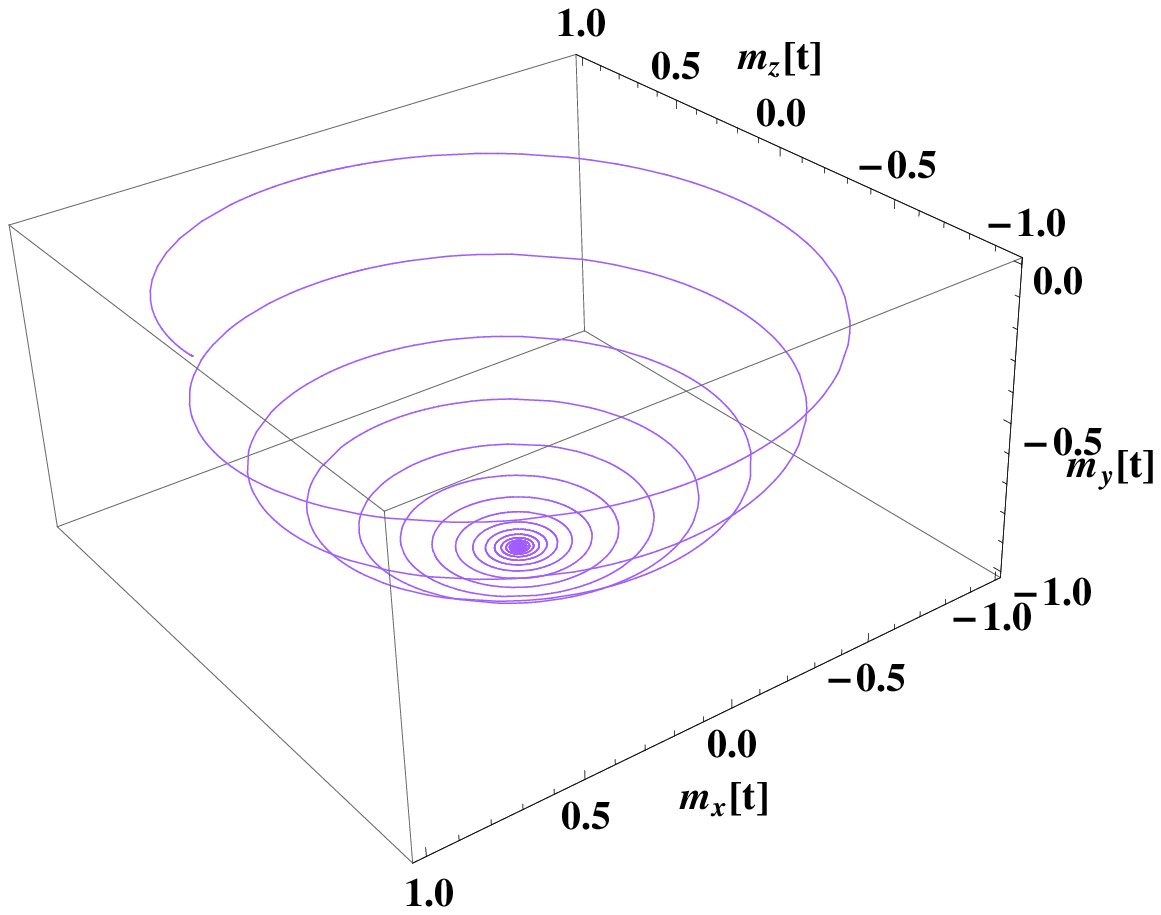}
\hspace{0.5cm}
\includegraphics[scale=0.46]{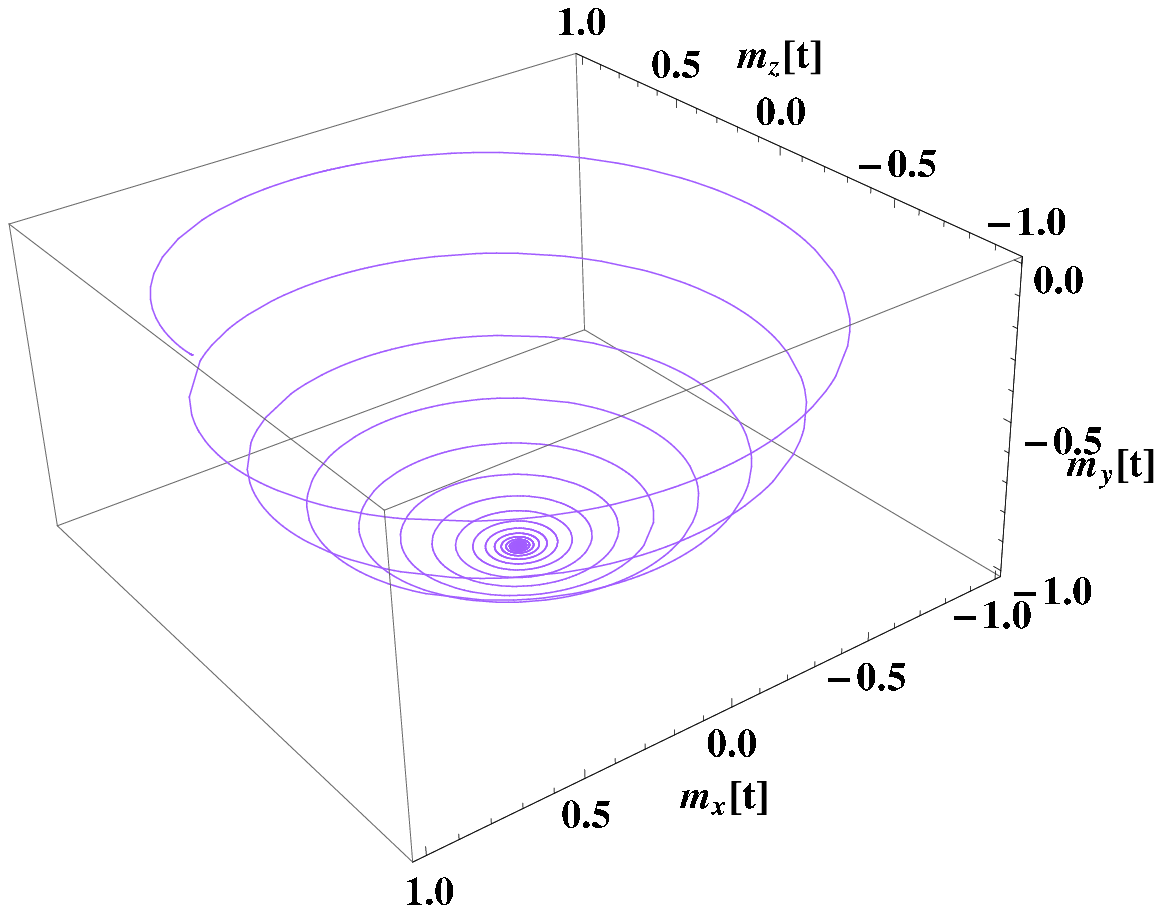}
}
\caption{(First four plots) The time evolution of the normalized components of 
magnetization with
initial angle of misalignment $\theta = 0.1\pi$. Here $\gamma_1 = 0.2$ with
$\gamma_2 = 45$ and $-45$ respectively in left and right for current biasing 
$I = 0.25$ mA in the top panels and $0.252$ mA for the bottom panels keeping
$B_0 = 0.1$ constant. The corresponding parametric graphs
representing the behaviour of the magnetization are shown in the last four 
plots.}
\label{fig3}
\end{figure*}

To investigate the magnetization dynamics and switching behaviour 
quantitatively, we have solved the full LLGS equations (\ref{eq6a}, \ref{eq6b},
\ref{eq6c}) of the F$|$S$|$F system using numerical simulation. The 
magnetization dynamics and switching behaviour of our system are investigated 
based on the above mentioned parameters, initially for very weak damping 
($\alpha\ll1$) and then for strong damping (up to $\alpha = 0.5$) with a very 
small angle of misalignment $\theta = 0.1\pi$ and $\gamma_1 = 2\gamma_2 = 0.2$.
Furthermore, to solve the equations (\ref{eq6a}, \ref{eq6b}, \ref{eq6c}) 
numerically the time coordinate has been normalized to $\tau = \gamma t /M_0 $,
where $M_0$ is the magnitude of the magnetization. Few of the corresponding 
numerical solutions are shown in Fig.\ref{fig2} for two  different current 
biasing in the first four plots. The rest plots in the figure show the 
corresponding parametric graphs of the time evolution of 
the magnetization components. The plots in left panels show the weak damping 
regime with Gilbert's damping parameter $\alpha = 0.05$ for two different 
choices of current biasing $0.1$ mA and $0.252$ mA respectively from top 
to bottom. While the damping is considered to be strong with $\alpha = 0.5$ in 
the plots of the right panels for the respective current biasings. It is 
seen that, the magnetization components show quite different behaviours. The 
components $m_{x}$ and $m_{z}$ display oscillating decay until they vanish 
completely, while on the other hand the component $m_{y}$ saturates with the 
increasing value of $\tau$. It is to be noted that, the qualitative behaviour 
of the components of magnetization is similar for different damping parameters.
But the quantitative difference is that, in strong Gilbert's damping regime, the
oscillation of the magnetization components $m_{x}$ and $m_{y}$ die out faster 
in time scale, while the component $m_{z}$ saturates too rapidly as seen from 
the right panels of Fig.\ref{fig2}. It is also seen that in strong damping, 
the reversal of magnetization components $m_y$ and $m_z$ does not occur for a 
current biasing $I = 0.252$ mA,  contrary to the case for the small damping. 
This result indicates that, it is possible to generate current induced 
magnetization reversal of a triplet superconducting ferromagnet in a F$|$S$|$F 
spin valve setup shown in Fig.\ref{fig1} by means of current biasing under
weak damping condition.

It is also our interest to see what happens when the parameter $\gamma_2$ is 
increased. To check the influence of $\gamma_2$ on switching mechanism, 
we investigated the behaviour of the magnetization components for both 
positive and negative values of $\gamma_2$ keeping the damping parameter and 
$\gamma_1$ fixed for currents respectively of $I = 0.25$ mA and $0.252$ mA.
It is found that the switching time $\tau$ gets delayed for $\gamma_2 = - 45,$ 
while on the other hand we observed a more rapid switching for $\gamma_2 = 45$. 
Moreover, the $x$ and $y$ components show more rapid oscillation for $\gamma_2 
= 45$  than for $\gamma_2 = -45$. The components of magnetization under this 
condition are shown in Fig.\ref{fig3} with the parametric graphs. This 
result suggests that magnetization reversal is dependent on strong coupling 
parameter and the switching of a system is more rapid for positive coupling
then that for negative coupling parameter as seen.

\begin{figure*}[hbt]
\centerline
\centerline{
\includegraphics[scale=0.42]{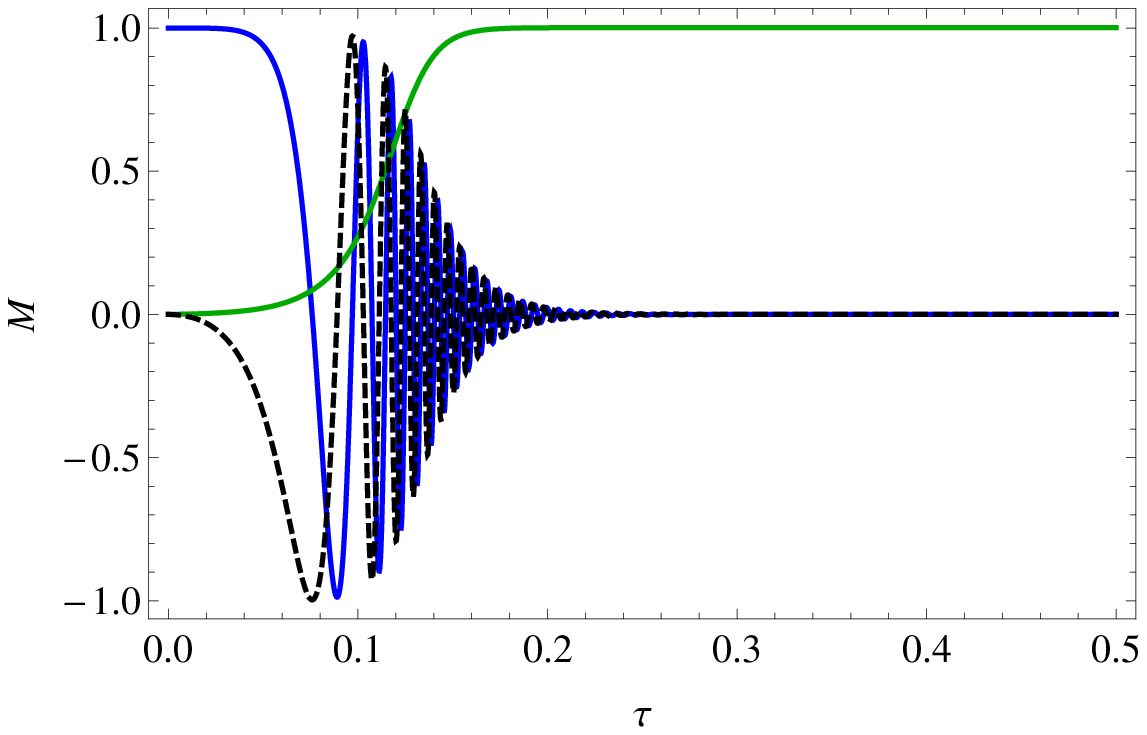}
\hspace{0.3cm}
\includegraphics[scale=0.42]{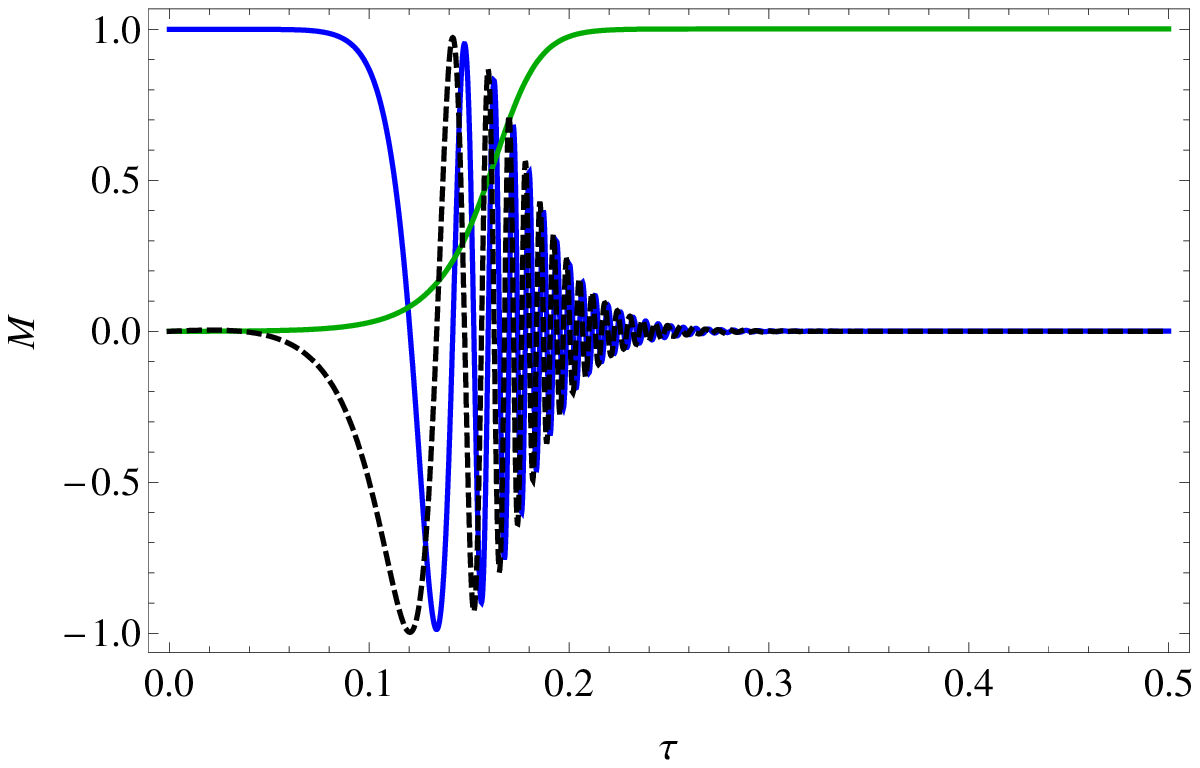}
\hspace{0.3cm}
\includegraphics[scale=0.42]{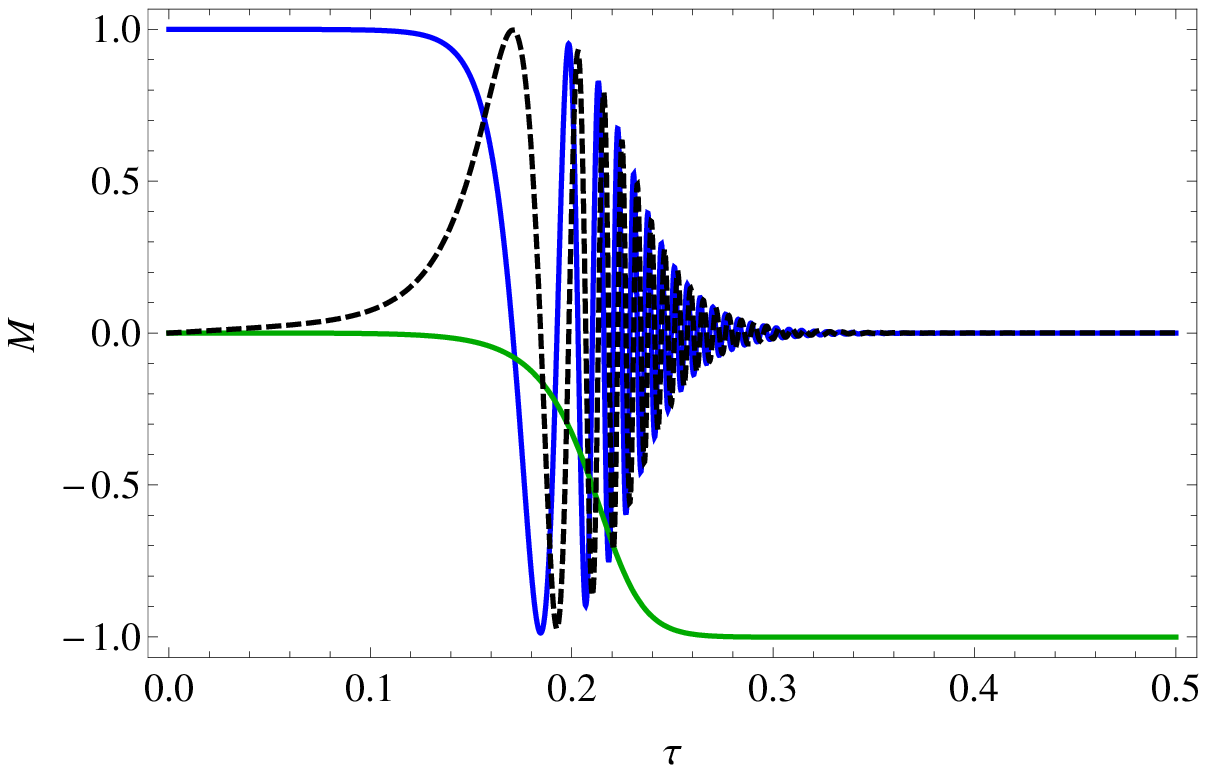}
}
\centerline{
\includegraphics[scale=0.42]{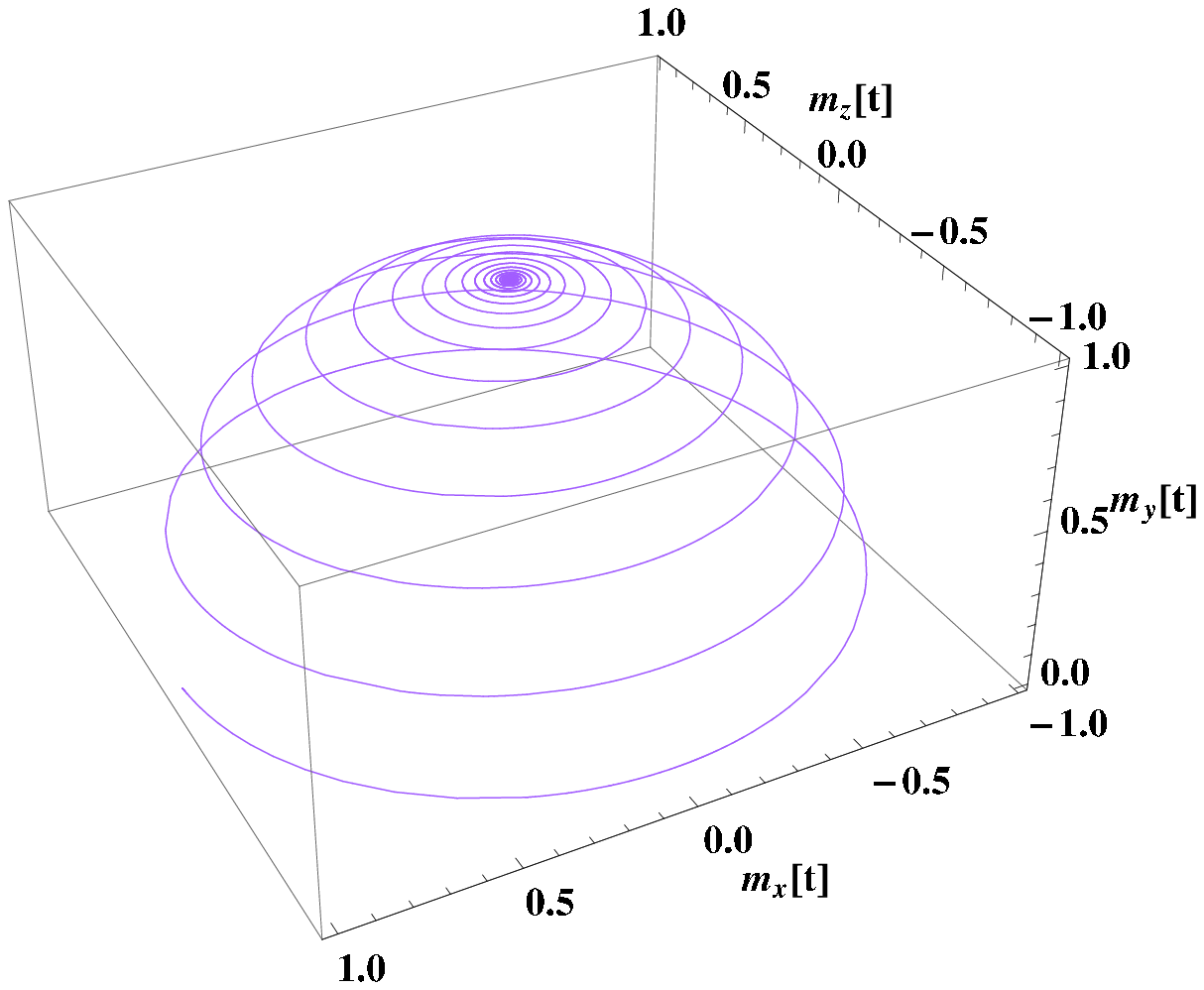}
\hspace{0.3cm}
\includegraphics[scale=0.40]{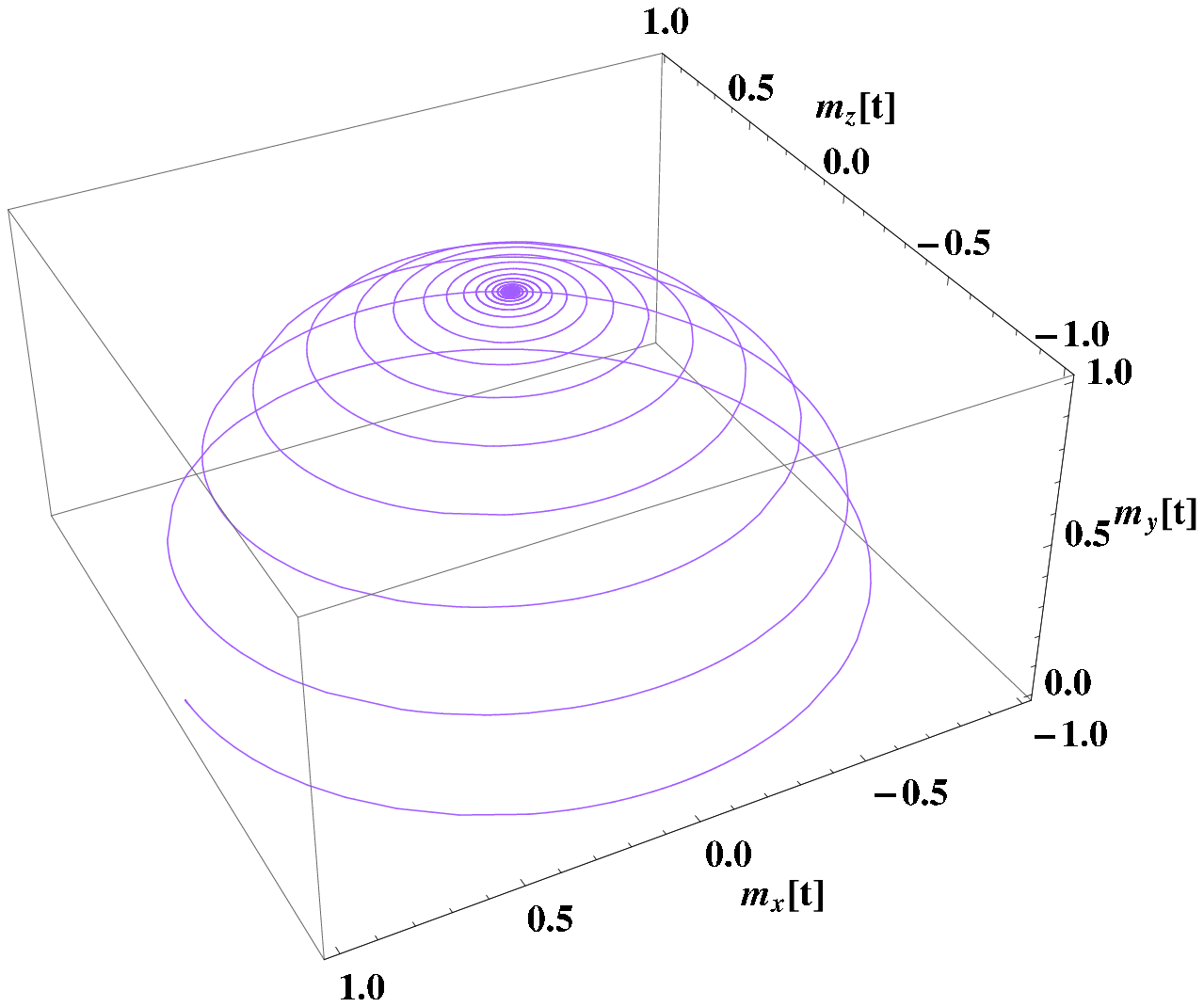}
\hspace{0.3cm}
\includegraphics[scale=0.44]{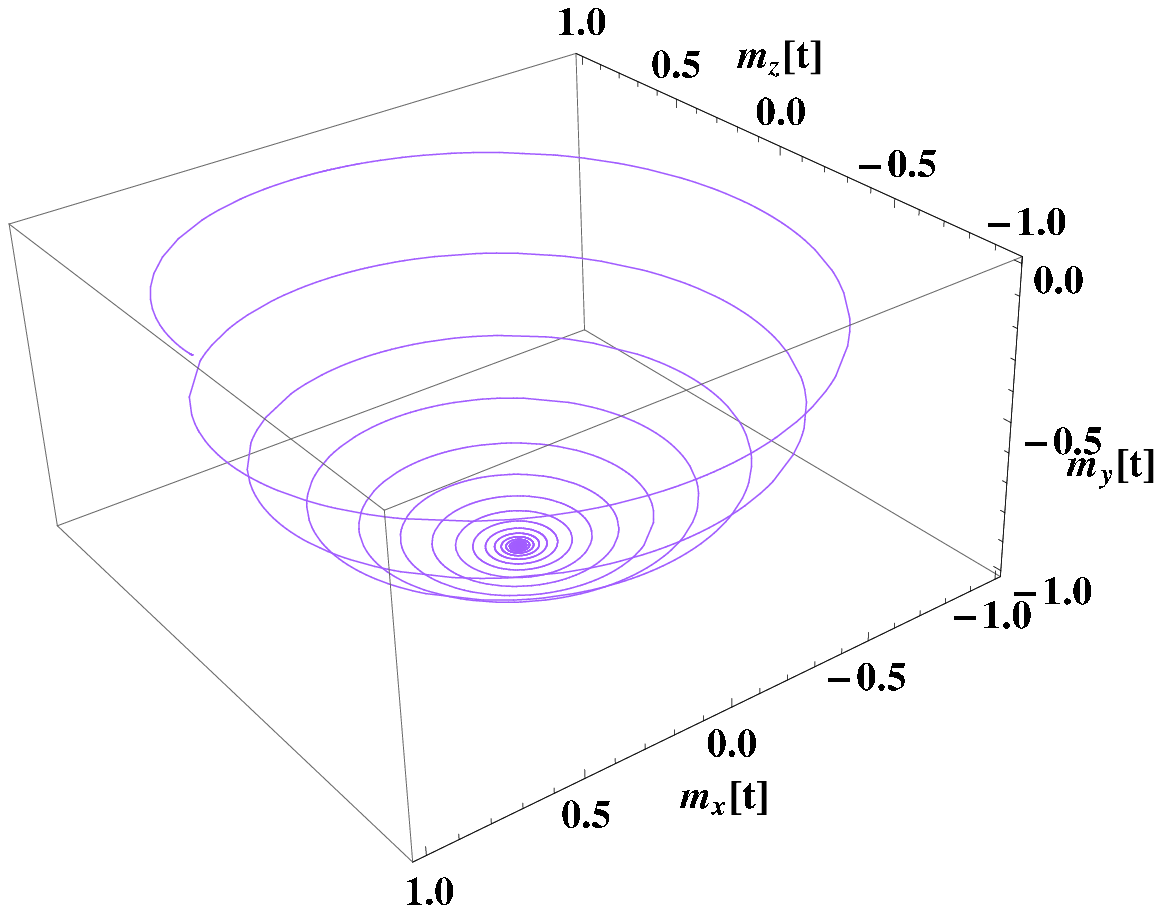}
}
\caption{(Top three plots) The time evolution of the normalized components of 
magnetization 
with initial angle of misalignment $\theta = 0.1 \pi$ for a magnetic induction 
$B_0 = 1.0$ with $\gamma_1 = 2\gamma_2 = 0.2$ and for weak damping $\alpha = 
0.05$. The plots in left and right depicted the magnetization dynamics for a 
current biasing of 0.1 mA and 1.65 mA respectively, while the plot in the 
middle is for a current biasing of 1.5 mA. The  
corresponding parametric graphs representing the behaviour of the magnetization
are shown in the bottom three plots.}
\label{fig4}
\end{figure*}

\begin{figure}[hbt]
\centerline
\centerline{\hspace{-0.4cm}
\includegraphics[scale=0.4]{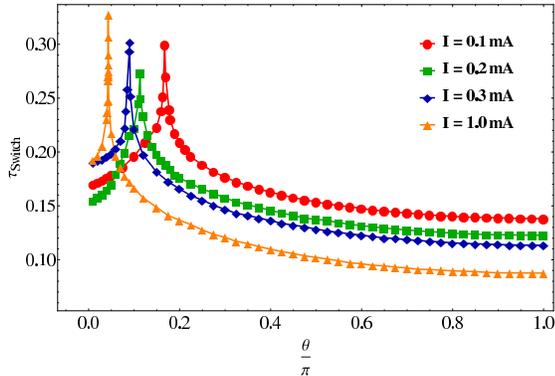}
}
\caption{Switching time and its dependence on the spin valve configuration for
different current biasings.}
\label{fig5}
\end{figure}

Our one more interest here is to check the influence of $B_0$ on switching. To 
investigate this we have plotted the magnetization components for a higher 
value of $B_0 = 1.0$ in Fig.\ref{fig4}. It is seen that under this situation 
the switching does not even occurs for a current biasing of $1.5$ mA as seen 
from the middle panel. In this configuration the reversal of the magnetization 
components occurs at a current biasing of $1.65$ mA as seen from right panel 
of Fig.\ref{fig4}. This suggest that the magnetization switching condition 
can also be controlled by magnetic induction.

It is to be noted from the Fig.\ref{fig2} and Fig.\ref{fig3} that for the 
weak damping but with higher current biasing, oscillations and switching for 
reversal of respective components of magnetization are delayed by some 
factors. In view of this result, it is also important to see explicitly what 
happens to the spin valve if configuration is changed and what influence the 
spin valve configuration has on the switching time $\tau_{switch}$? 

To answer these two questions, we have studied the switching time 
$\tau_{switch}$ as a function of $\theta$ representing the angles of
misalignments for four different current biasings keeping the damping
factor $\alpha = 0.05$ as shown in the Fig.\ref{fig5}. It should be noted 
that, the switching time of a  magnetization component is defined as the time 
required by the component to attain numerically the $0.975$ times of its 
saturated value \cite{linder1}. One of the important results of this study is 
that, for the increasing angle of misalignment, more rapid switching of 
corresponding components of magnetization occurs with the increasing value of 
the current bias. From the Fig.\ref{fig5} it is also seen that, for all 
current biasing the 
switching time shows monotonic increase with a sharp peak staring from the
zero angle of misalignment, providing the most delayed magnetic spin valve 
configuration at a particular current. The angle of misalignment of this most 
delayed configuration decreases with increasing value of the biasing current.
It is interesting to note that, the maximum switching time for a particular
angle of misalignment increases with increasing value of current biasing 
except for the case of $0.2$ mA current, at which it is lowest. This 
particular behaviour at $0.2$ mA current indicates that in the range of 
smaller angle of misalignment ($\theta \le 0.05 \pi$), $0.2$ mA is the 
optimum value of biasing current among all for the magnetic spin valve. The 
data of these results are summarized in Table \ref{tab1}. These results as a 
whole clearly signify that, switching is highly dependent on magnetic 
configuration in association with the biasing current: switching occurs 
swiftly at higher angle of misalignment with higher value of current
bias. This suggests that, the configuration near the anti-parallel ($\theta = 
\pi$) offers rapid switching then the parallel ($\theta = 0$) for all 
the current biasing, and for higher current, the peak position shifted towards 
the parallel configuration lowering the switching time just after the peak. 
This is quite obvious as the STT becomes stronger in this case.
\begin{center}
\begin{table}[ht]
\caption{\label{tab1} Maximum switching time ($\tau_{switch}$) and
corresponding misalignment angle ($\theta$) for different biasing currents in
low damping with $\alpha = 0.05$.}\vspace{0.3cm}
\begin{tabular}{||c|c|c||}\hline\hline
I (in mA) & $\theta$ & $\tau_{switch}$ (in Sec) \\\hline
0.1 &  0.168$\pi$  &  0.2995 \\
0.2 &  0.114$\pi$  &  0.2734 \\
0.3 &  0.091$\pi$  &  0.3022 \\
1.0 &  0.044$\pi$  &  0.3280 \\
\hline\hline
\end{tabular}
\end{table}
\end{center}
\begin{figure*}[hbt]
\centerline
\centerline{
\includegraphics[scale=0.45]{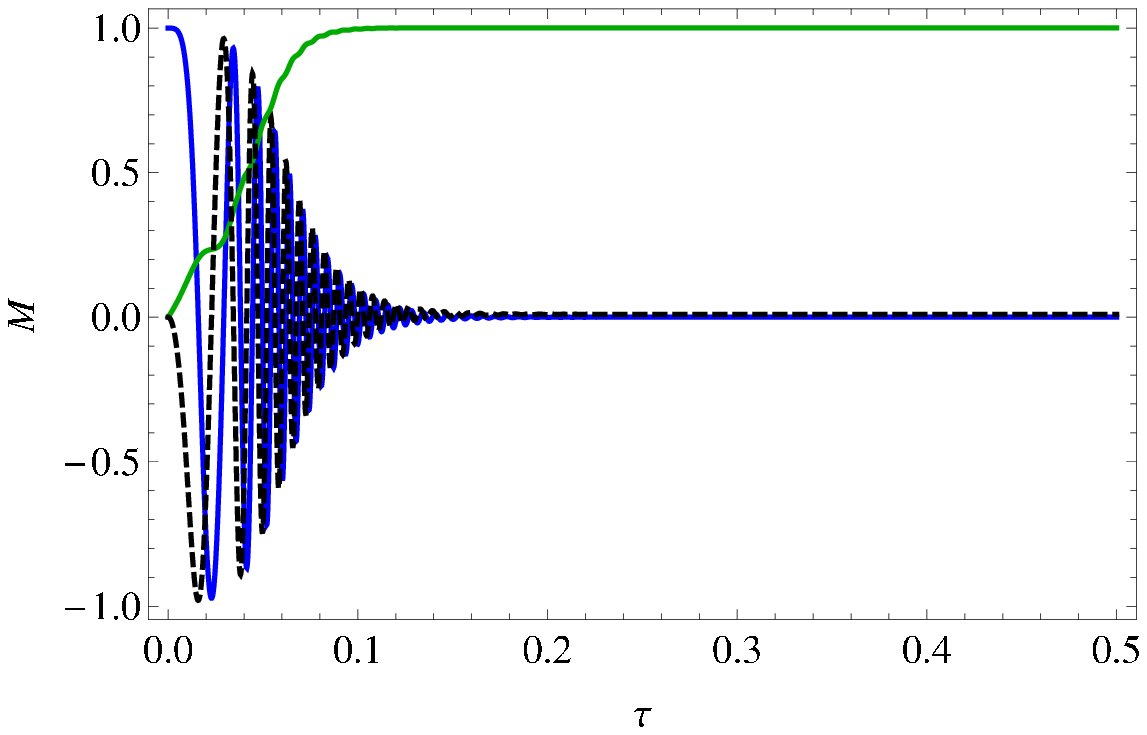}
\hspace{0.5cm}
\includegraphics[scale=0.45]{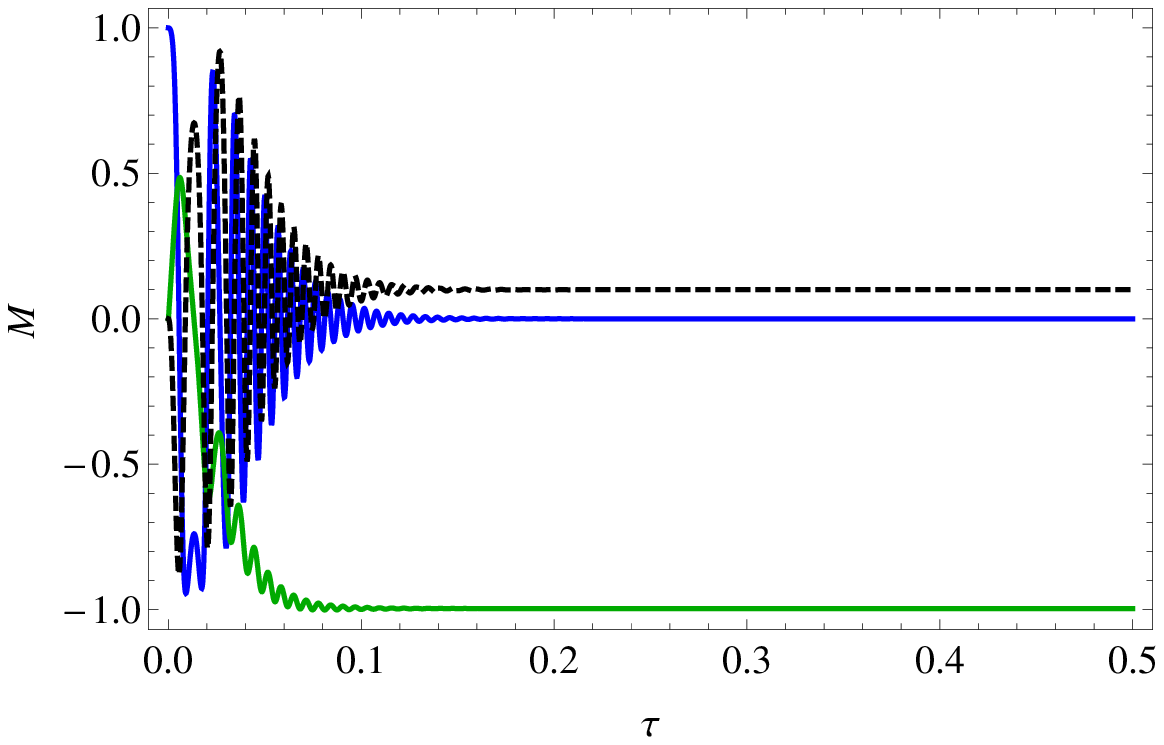}
\vspace{0.3cm}}
\centerline{\hspace{0.05cm}
\includegraphics[scale=0.45]{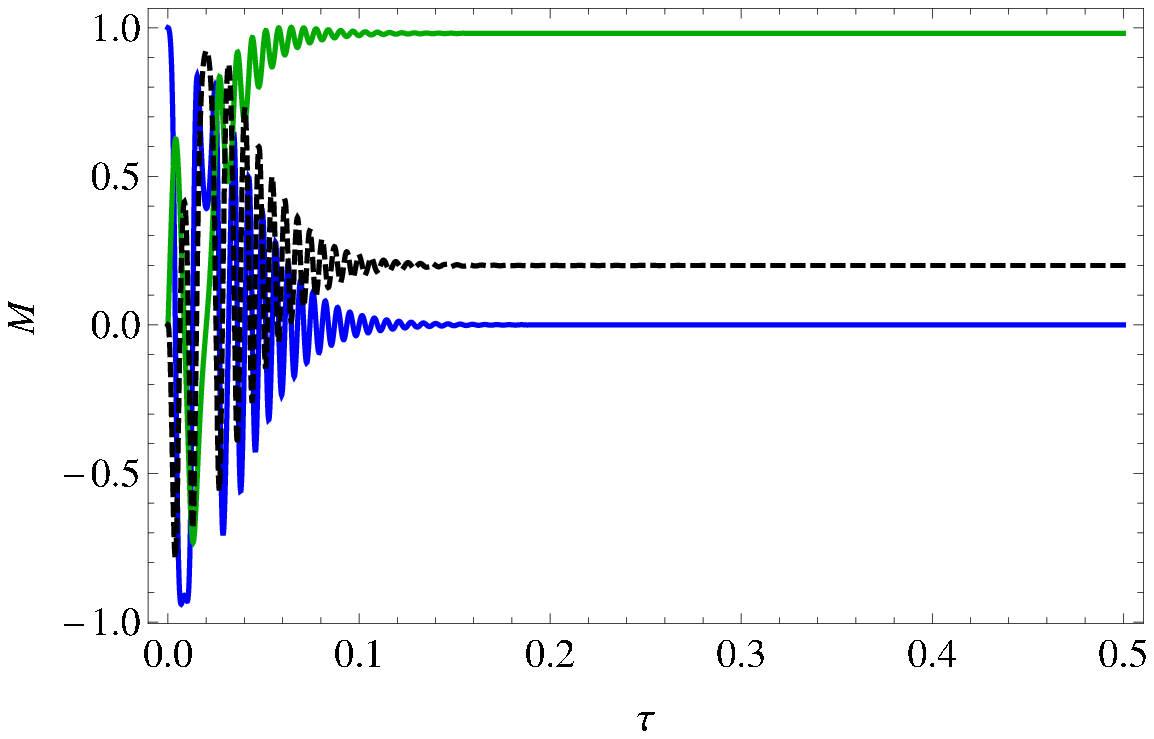}
\hspace{0.4cm}
\includegraphics[scale=0.456]{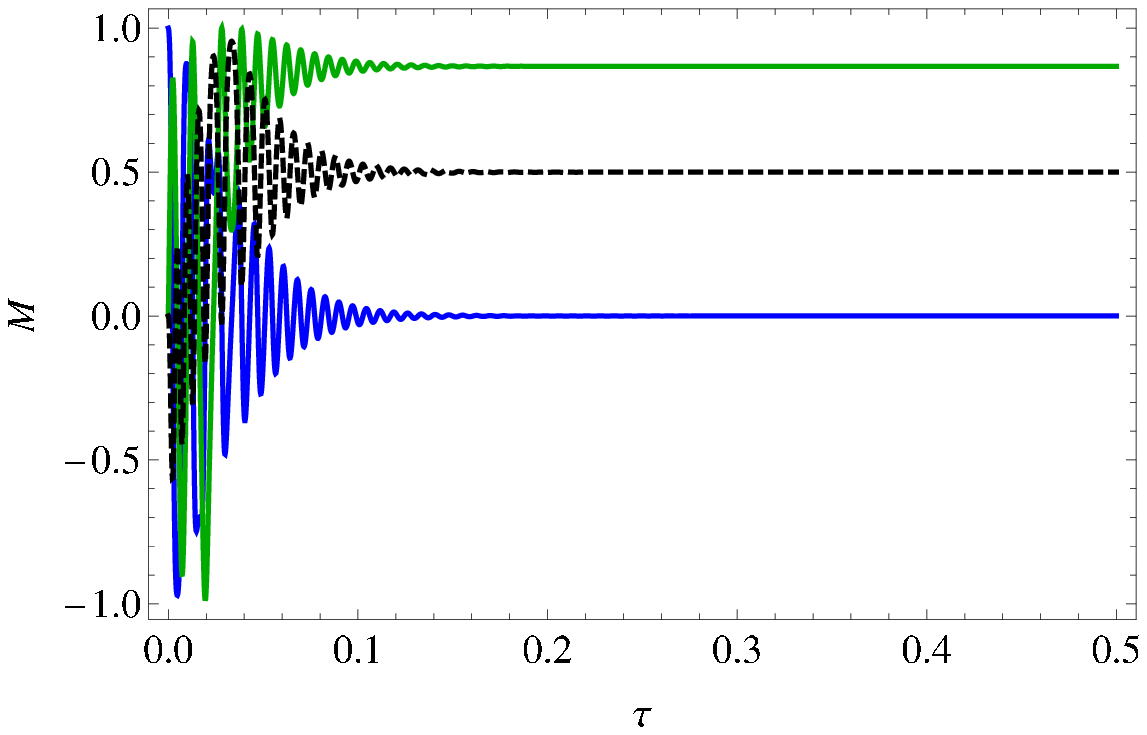}
}
\centerline{
\includegraphics[scale=0.45]{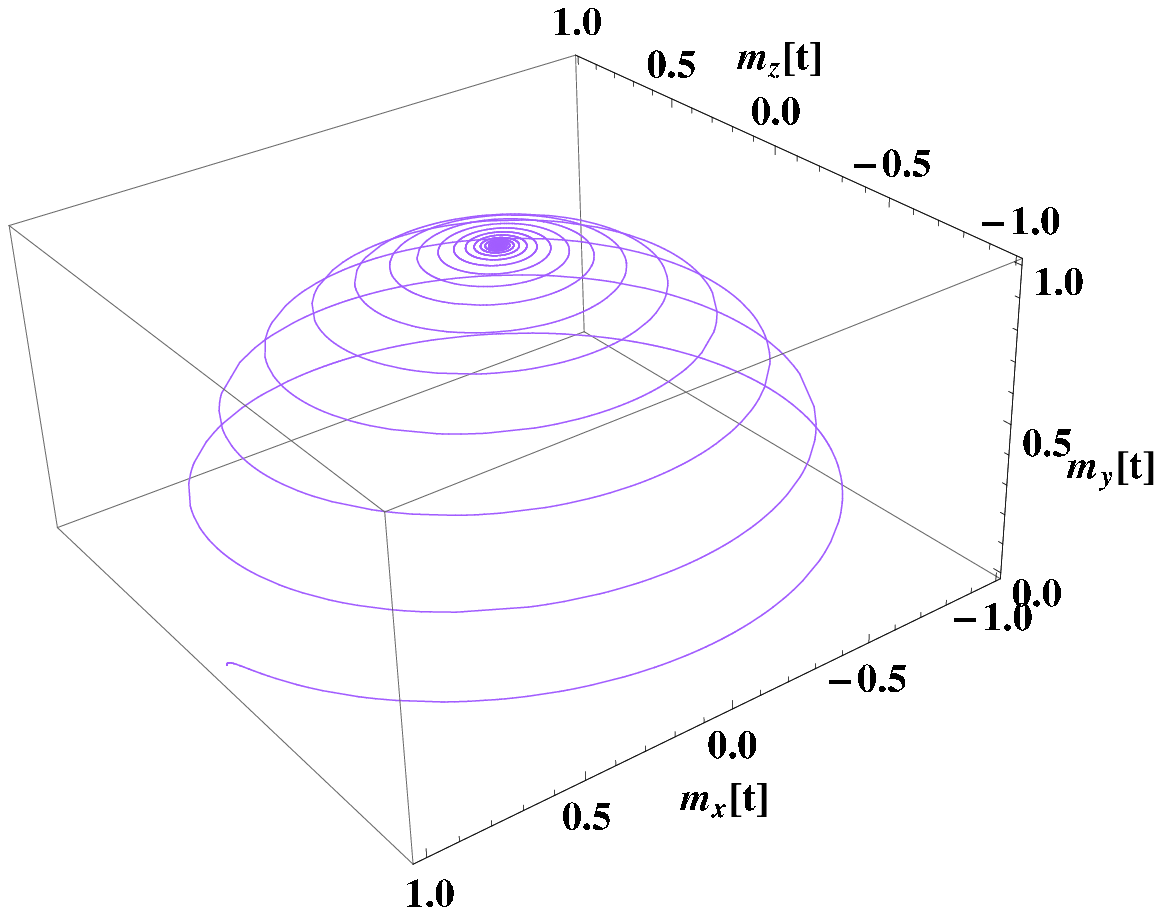}
\hspace{0.5cm}
\includegraphics[scale=0.45]{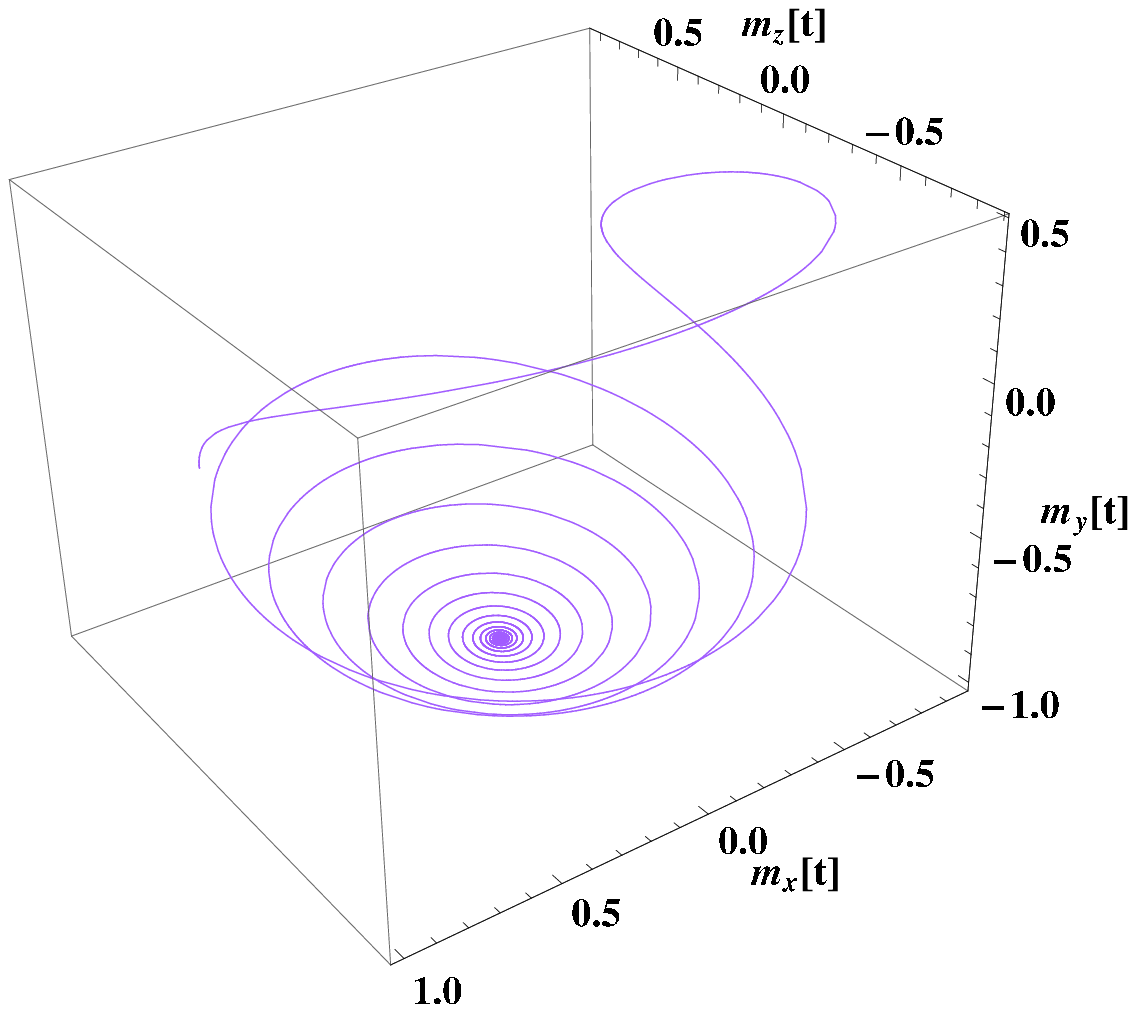}
\vspace{0.3cm}}
\centerline{
\includegraphics[scale=0.45]{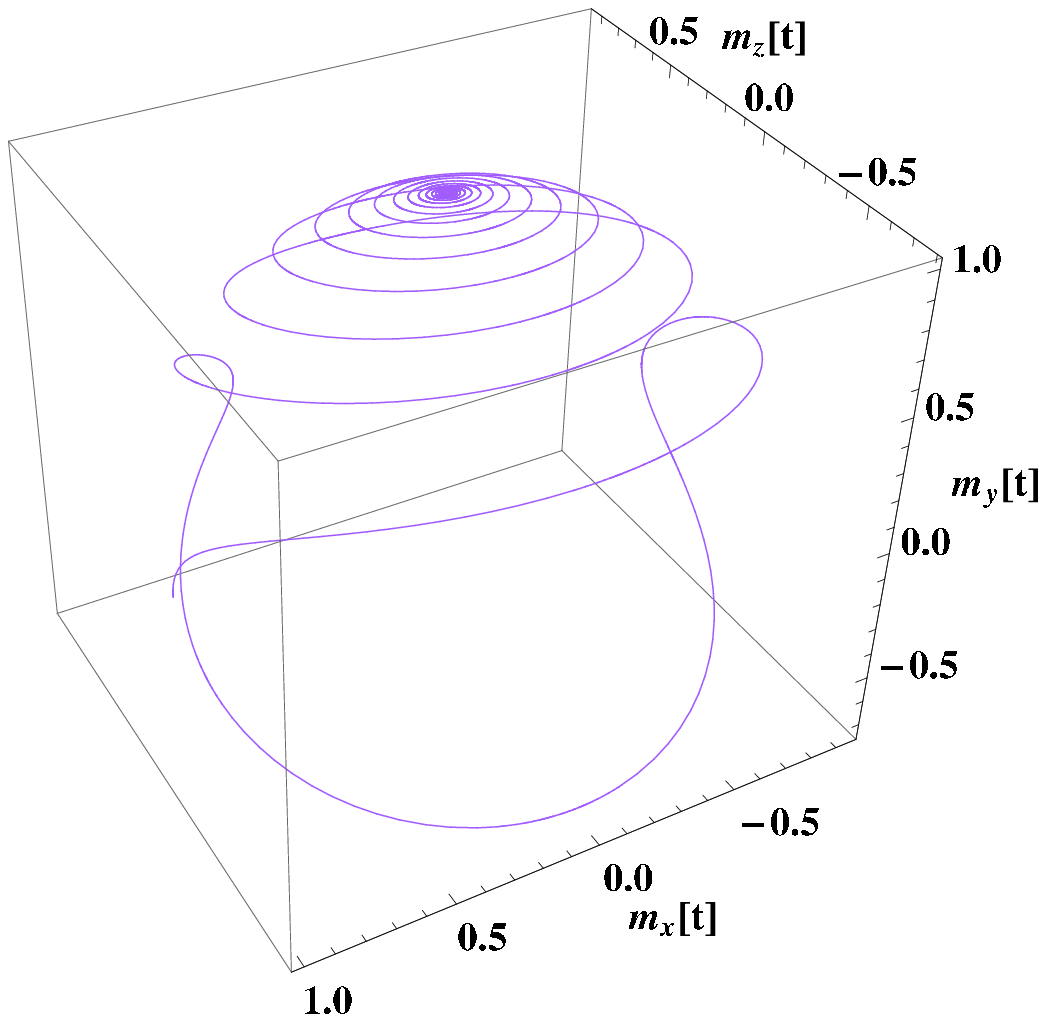}
\hspace{0.5cm}
\includegraphics[scale=0.45]{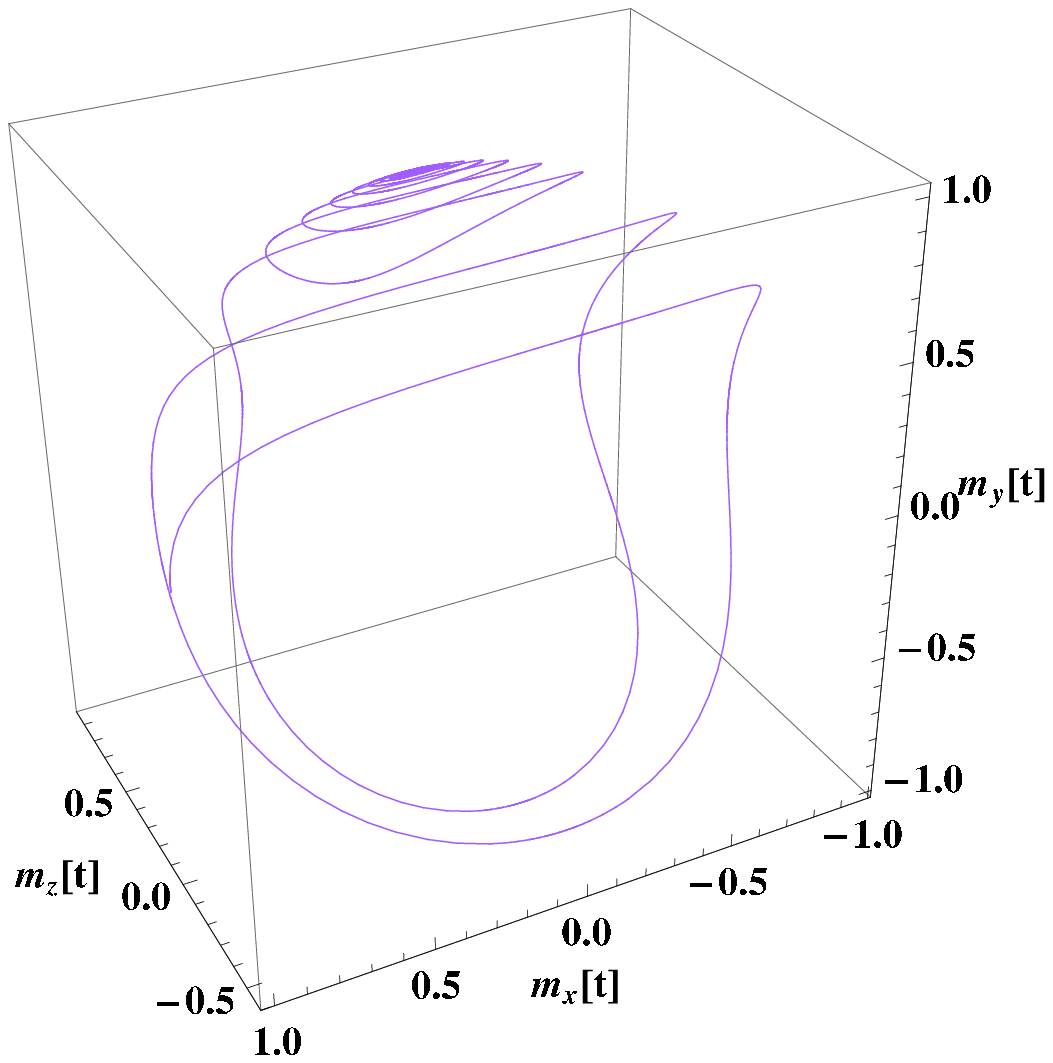}
}
\caption{(First four plots) The time evolution of normalized components of 
magnetization with 
initial angle of misalignment $\theta = 0.1 \pi$ with $\gamma_1 
= 2\gamma_2 = 0.2$ and for current biasing $I = 0.1$ mA for higher values 
of external magnetic field, viz., $B_0 = 10^2$ (top left), $B_0 = 10^3$ (top 
right), $B_0 = 2\times10^3$  (bottom left) and $B_0 = 5\times10^3$ (bottom 
right). The corresponding parametric graphs representing the 
behaviour of the magnetization are shown in the last four plots.}
\label{fig6}
\end{figure*}
Our final interest is to study the influence of higher magnetic field on the 
component of magnetization.  To analyze this we have plotted the magnetization 
components with $\tau$ for higher values of the magnetic field in 
Fig.\ref{fig6}, keeping the biased currents fixed at $0.1$ mA and 
$\gamma_1 =2\gamma_2 = 0.2$. It is seen that for $B_0 = 10^2$, the components 
of magnetization shows quite similar behaviour as seen earlier in Fig.\ref{fig2}. 
However, on the other hand as the magnetic field increases the 
components of magnetization shows quite irregular behaviour. For example, for 
$B_0 = 10^3$ as seen from the top right panel of the Fig.\ref{fig6}, the 
$m_y$ component suddenly reverses with a small initial fluctuations and then 
starts saturating after a oscillating decay period. This is due to the fact 
that, as $B_0$ becomes of the order of the anisotropy field, the components of 
magnetization behave quite differently. In this condition, the component
$m_y$ reverses and saturates, while $m_x$ and $m_z$ show an oscillating decay.
With further rise in $B_0$ makes the system more unstable in such a way that,
with increasing value of $B_0$, both $m_y$ and $m_z$ components gradually tend 
to behave almost similarly by retaining the original direction of the $m_y$ 
component as seen from the bottom panels of the first four plots in the 
Fig.\ref{fig6}. Because, with 
further rise in $B_0$, the magnetic field dominates over the anisotropy field. 
It can be easily visualized from the parametric graph shown in the bottom left 
of panel of the Fig.\ref{fig6}, where the motion takes place about the 
direction of magnetic field. The motion stabilizes itself for more higher 
values of $B_0$. We have found that, the influence of magnetic field 
as mentioned above is almost similar for the biasing current and hence 
higher value of magnetic field ($B_0 \ge 100$) eliminates the effect of biasing
current.     

\section{Summary}
In this work, we have investigated the current induced magnetization dynamics  
and magnetization switching in a superconducting ferromagnet sandwiched between 
two misaligned ferromagnetic layers with easy-axis anisotropy by numerically 
solving Landau-Lifshitz-Gilbert-Slonczewski's equation. For this purpose, we 
have used the modified form of the Ginzburg-Landau free energy functional for a 
triplet p-wave superconductor. We have demonstrated about the possibility of 
current induced magnetization switching for an experimentally realistic 
parameter set. It is observed that, for the realization of magnetization 
switching sufficient biased current and moderate field are suitable for the 
case of low Gilbert damping. Although, switching can be delayed for large 
damping, however such system can not be used because the system become highly 
unstable in such situation, which is unrealistic. It is also to be noted that 
switching is highly dependent on the strong coupling parameter and it is seen 
that positive value of that offers more rapid switching then that of negative. 
It is also seen that switching has a high magnetic configuration dependence. 
It shows a monotonic increase for both low and high current in very near to 
parallel configuration. The configuration near anti parallel offers more rapid 
switching than the parallel. Again, it can also be conclude that the dynamics 
is highly dependent and controlled by the magnetic field as it becomes of the 
order of the anisotropy field. As a concluding remark, the results indicate 
about the switching mechanism in F$|$S$|$F spin valve setup for an 
experimentally favourable parameter set, which may be utilized to bind 
superconductivity and spintronics \cite{linder2} together for making 
practical superconducting-spintronic devices.\\

\section*{Acknowledgments}
Authors are thankful to Professor Jacob Linder, Department of Physics, 
Norwegian University of Science and Technology, N-7491 Trondheim, Norway for 
his very helpful comment during communication, which leads to a considerable 
improvement of the work.

\end{document}